# THE HST QUASAR ABSORPTION LINE KEY PROJECT
# VII. ABSORPTION SYSTEMS at $z_{\rm abs} \leq 1.3$[1]

John N. Bahcall[2], Jacqueline Bergeron[3], Alec Boksenberg[4], George F. Hartig[5], Buell T. Jannuzi[2,11], Sofia Kirhakos[2], W. L. W. Sargent[6], Blair D. Savage[7], Donald P. Schneider[2,12], David A. Turnshek[8], Ray J. Weymann[9], and Arthur M. Wolfe[10]

[2]Institute for Advanced Study, School of Natural Sciences, Princeton, NJ 08540

[3]Institut d'Astrophysique, 98 bis Boulevard Arago, F-75014 Paris, France

[4]Royal Greenwich Observatory, Madingley Road, Cambridge CB3 0EZ, England

[5]Space Telescope Science Institute, 3700 San Martin Drive, Baltimore, MD 21218

[6]Robinson Laboratory 105-24, California Institute of Technology, Pasadena, CA 91125

[7]Department of Astronomy, University of Wisconsin, 475 N. Charter Street, Madison, WI 53706

[8]Department of Physics & Astronomy, University of Pittsburgh, Pittsburgh, PA 15260

[9]The Observatories of the Carnegie Institution of Washington, 813 Santa Barbara Street, Pasadena, CA 91101

[10]Center for Astrophysics & Space Sciences, C011, University of California San Diego, La Jolla, CA 92093

[11]Hubble Fellow

[12]Department of Astronomy, The Pennsylvania State University, University Park, Pennsylvania 16802

## ABSTRACT

We present evidence that clumps of Ly-$\alpha$ lines are physically associated with about half of the extensive metal-line systems (absorption systems with four or more observed metal-line species) found in this paper, demonstrate that all four Lyman-limit systems discussed here correspond to extensive metal-line absorption systems, and present an extraordinary pair of extensive metal-line absorption systems within 2000 km/s of each other at $z = 0.95$ that are probably an early manifestation of large scale structure. These results are obtained using ultraviolet spectra, taken with the higher-resolution gratings of the Faint Object Spectrograph of the Hubble Space Telescope, for four quasars with emission-line redshifts between 1.0 and 1.3. We also determine the evolution of Ly-$\alpha$ absorption lines at redshifts less than 1.3 by combining the results for 13 smaller redshift quasars discussed in Paper I of this series with the 4 moderate redshift quasars analyzed in the present paper.

Absorption lines were selected, measured, and identified algorithmically using software tested by Monte Carlo simulations. A total of 291 absorption lines, all with a statistical significance above a specified high threshold level, were selected and measured. A total of 145 lines are identified as extragalactic Ly-$\alpha$ absorption lines. Ten of the Ly-$\alpha$ absorption lines are found at the same redshifts as metal-line systems. Monte Carlo simulations with pseudo-C IV or O VI doublets were carried out to determine the probability that a pair of absorption lines might accidentally have the appropriate separation to be identified as either a C IV or as an O VI absorption doublet. The average number of pseudo C IV doublets found in the real (observed) spectra varies from 0.05 to 2.4 per spectrum within the Ly-$\alpha$ forest and is negligible outside the Ly-$\alpha$ forest.

For $z_{\text{abs}} \leq 1.3$, the density of Ly-$\alpha$ lines with equivalent widths greater than 0.24 Å is adequately fit by $(dN/dz) = (dN/dz)_0 \cdot (1+z)^\gamma$ with $(dN/dz)_0 = 24.3 \pm 6.6$ Ly-$\alpha$ lines per unit redshift, and $\gamma = 0.58 \pm 0.50$ (1-$\sigma$ uncertainties). This rate of evolution at low redshifts is less than the evolutionary rate inferred from several different ground-based data samples that pertain to high redshifts, although neither the available HST data nor the ground-based data are sufficiently extensive to establish whether this change occurs abruptly or gradually.

The four Lyman-limit systems that are present in the spectra analyzed here all correspond to extensive metal-line systems. This result provides further circumstantial evidence that many Lyman-limit systems (like many metal-line absorption systems) are associated with galaxies.

Eight extensive metal-line systems with between five and fifteen strong metal lines are identified. An approximate estimate for the frequency of such systems is $dN/dz \simeq 2.5(1+z)^{0.5}$ systems per unit redshift or $dN/dz \simeq 2.0(1+z)^{1.0}$ systems per unit redshift.

About half of the extensive metal-line systems are accompanied by clumps of neighboring (in redshift space) Ly-$\alpha$ absorption lines, corresponding to velocity dispersions of 600 km s$^{-1}$ to 1400 km s$^{-1}$. In addition, two of the extensive metal systems, found in the spectrum of PKS 0122$-$00 at $z = 0.9667$ and $z = 0.9531$, are probably physically associated since they are separated by only 2000 km s$^{-1}$. We suggest that the metal-line systems with associated clumps of Ly-$\alpha$ lines and the linked pair of metal-line systems seen in the spectrum of PKS 0122$-$00, may correspond to clusters, or possibly superclusters, of galaxies.

The observed gaseous structures at redshifts of 0.5 to 1.0 with velocity dispersions of $6 \times 10^2$ km s$^{-1}$ to $1.4 \times 10^3$ km s$^{-1}$ (or velocity spans of $1.2 \times 10^3$ km s$^{-1}$ to $3 \times 10^3$ km s$^{-1}$) constitute a constraint on cosmological models of structure formation. The local column density (number density times radius squared) for the clumps of Ly-$\alpha$ absorptions and metal-line systems is $10^{-4}$ Mpc$^{-1}$.

The clumps of Ly-$\alpha$ absorption lines clustered about metal-line systems and the inferred rate of evolution of low and moderate redshift Ly-$\alpha$ absorption lines more nearly resemble the properties of galaxies and of metal-containing absorption line systems than they do the properties of the high-redshift Ly-$\alpha$ forest lines. These results are consistent with two different populations of Ly-$\alpha$ absorption lines, with type 1 being closely associated with galaxies and evolving relatively slowly and type 2 being relatively unclustered, evolving more rapidly, and dominating the observations at large redshift.

*Subject headings:* cosmology: observations — quasars: Ton 153, PKS 0122−00, PG 1352+011, PG 1634+706

## 1. Introduction

Observations with ground-based telescopes have determined many of the characteristics of Ly-$\alpha$ absorption systems at redshifts larger than $z = 1.7$ (e.g., Murdoch et al. 1986; Carswell 1988; Hunstead 1988; Rees 1988; Sargent 1988; Lu, Wolfe, & Turnshek 1991; Press, Rybicki, & Schneider 1993; Bechtold 1994) and observations with the Hubble Space Telescope (Bahcall et al. 1991a, 1992, 1993b; Morris et al. 1991; Bahcall et al. 1993a, hereafter Paper I) have established some of the characteristics of Ly-$\alpha$ absorption systems at small redshifts ($z \leq 0.5$). The principal purpose of this paper is to present the data and the analyses that provide the first description of the characteristics of Ly-$\alpha$ absorption systems at the intermediate redshifts: $0.6 < z < 1.3$. In the process, we have also discovered some important facts regarding 'extensive metal-line absorption systems,' a term which we will use in the following to mean an absorption system in which four or more metallic ions are detected.

Much of the background material necessary to understand this paper has been given previously in Paper I, which concentrated on the analysis of the spectra of small-redshift quasars, in Schneider et al. (1993, hereafter Paper II), which described our original algorithmic methods for calibrating the data and for selecting the absorption lines, and in Savage et al. (1993b, hereafter Paper III), which determined the characteristics of the Galactic interstellar absorption lines. We will summarize only briefly here those topics that are discussed more fully in these previous Key Project papers and will concentrate in the present paper on describing improvements in the previous methods and on the new results for the intermediate-redshift absorption line systems.

We analyze in this paper the combined sample of previously-studied small-redshift quasars and the four intermediate redshift quasars first discussed here. The combined sample covers the redshift range $0 \leq z_{\rm abs} \leq 1.3$.

The paper is organized as follows: summary of the observations and a brief description of how the data are calibrated (§ 2.), description of the current algorithms for selecting absorption lines and measuring their properties (§ 3.), a summary of the current algorithms for identifying absorption lines (§ 4., software *ZSEARCH*), searches for simulated doublets (§ 5.), line identifications for Ton 153, PKS 0122−00, PG 1352+011, and PG 1634+706, with comments on the identifications of especially interesting individual absorption systems (§ 6.), an analysis of the evolution of Ly-$\alpha$ absorption lines (§ 7.), and a discussion of the clumping of Ly-$\alpha$ absorption lines near extensive metal-line systems (§ 8.). We summarize and discuss our main results in (§ 9.). Absolute dimensions in this paper are calculated assuming $H_0 = 100$ km s$^{-1}$ Mpc$^{-1}$ and $q_0 = 0.5$.

## 2. Observations

All observations were made using the Faint Object Spectrograph (FOS) of the Hubble Space Telescope. The observing and data calibration procedures are described in detail in Paper II; a description of the FOS is given by Ford & Hartig (1990). The observations analyzed in this paper are listed in Table 1. The FOS has two detectors, denoted as "Blue" and "Red", used for observations below and above 1600 Å, respectively. We used two higher resolution gratings (G190H and G270H, $R = 1300$) to observe between 1600–3270 Å and a low resolution grating (G160L, $R \approx 180$) to observe between 1150–2400 Å. The full widths at half maximum (FWHM) of unresolved lines observed with the different gratings are: 1.5 Å (G190H), 2.0 Å (G270H), and 9.4 Å (G160L). Table 1 also includes basic information about the four quasars that are investigated in this study, including their Véron-Cetty & Véron (1991) catalogued redshifts and $g$-magnitudes (Thuan & Gunn 1976) as reported in Kirhakos et al. (1994). (For low-redshift quasars, $g \approx V - 0.1$.) In addition we have listed the J2000 coordinates provided by the Space Telescope Science Institute (STScI). The coordinates were measured from the digitized version of the "Quick $V$" Survey plates (Epoch 1982) described in Lasker et al. (1990) using STScI's Guide Star Selection System Astrometric Support Package (GASP) and should be accurate to about 1″.

Corrections for variations in sensitivity of the photocathodes and diodes of the FOS detectors are made from flat fields generated from observations of stars. For all of our observations, we have used the flat field generated from calibration observations closest in time to the quasar observations. The flat-field structure is particularly strong in the central portions of the G190H spectra. Residual errors in the flat-fielding may dominate the statistical noise in some portions of the spectra and can create spurious weak features (cf. Paper II and § 3.).

Table 2 lists observed features that are strong enough to be in the "complete sample" of lines (defined below), but are recognized by comparison with observations of standard stars to be residual flat-field features (cf. Jannuzi & Hartig 1994).

We have placed all the higher-resolution observations for a given object (different gratings, e.g. G190H and G270H spectra) on a common wavelength scale by requiring that the strong, singly-ionized ISM absorption lines are at rest (see Paper II for details). In Table 3, we list for each spectrum the wavelength zeropoint-offsets that were *added* to the reduced spectra. No correction could be made for our observations of PKS 0122−00 because the necessary Galactic absorption lines were not detectable. In the final column of Table 3 we list the small additional offset that should be added to our spectra and line lists to place them in the heliocentric rest frame. These later offsets are based on work described in Paper III and in Lockman & Savage (1994).

Figure 1 shows the calibrated HST data used to construct the database presented in this paper. The calibrated data consist of two arrays: the flux and the $1\sigma$ uncertainty in the flux as a function of wavelength. The continua were constructed by fitting cubic splines to a collection of points that represents our best estimate of the continuum flux. The continuum for each object is shown as a dotted line in Figure 1. In addition to defining the continuum, the fitted curves attempt to reproduce the emission line profiles and any broad absorption profiles that may be present. (See Paper II for an expanded discussion of the continuum fitting procedure.) The spectra and the associated errors are divided by the continua to produce normalized arrays for our analysis program. Figure 1 also displays for each source the adopted $4.5\sigma_{\rm det}$ minimum equivalent width limit as a function of wavelength that must be exceeded in order for an absorption line to be included in the complete sample (cf. Paper II and the last paragraph in § 3.). Features that are known to be caused by imperfect flat fielding are labeled FF in Figure 1.

## 3. Selection and Measurement of the Absorption Lines

This section gives a brief overview of the selection and measurements of the absorption lines; we present in Paper II a detailed description of the software and the algorithms. The notation for the line-fitting algorithms in this paper is the same as that used in Paper II. At the end of this section, we describe the procedure that is used in the present paper to measure the redshifts of the Lyman-limit Systems (LLSs).

A normalized spectrum for each of the quasars was created by dividing the calibrated data by the continuum fit. The fluxes in unresolved lines (absorption or emission) centered on each pixel in the normalized spectrum were calculated using the procedure described in Section 6 of Paper II. The Spectral Spread Functions (SSFs) in all of the Key Project observations have Gaussian profiles (see Figure 1 of Paper II). Each line is assigned a significance level, $SL$:

$$SL \ = \ \frac{|W|}{\overline{\sigma}(W)} \ , \qquad (1)$$

where $W$ is the observed equivalent width, and $\overline{\sigma}(W)$ is the $1\sigma$ error in the observed equivalent width of an unresolved line centered at that pixel. Note that $\overline{\sigma}(W)$ is calculated with the flux errors at the positions of strong absorption lines replaced by the errors calculated from the surrounding continuum points (see Paper II). The significance level differs from the more familiar definition of signal-to-noise ratio, $SNR = |W|/\sigma(W)$, because of the use of the *interpolated* error array for the calculation of $SL$. This SSF search procedure is iterated a number of times (nine times in the analysis described in this paper) so that features consisting of several closely spaced lines can be deblended into individual components.

At this point one constructs the function $\sigma_{\text{det}}(\lambda_{\text{obs}})$ from the array of $1\sigma$ errors ($\overline{\sigma}(W)$; see Figures 9 and 10 in Paper II). A preliminary line list is created using the features whose equivalent widths exceed a selected $SL$ threshold ($W > C_{SL}\sigma_{\text{det}}(\lambda_{\text{obs}})$; $C_{SL} = 3.0$ for the preliminary line lists in this paper). Each of these lines is fitted with a variable-width Gaussian profile; this allows one to characterize the properties of resolved lines. Note that this procedure only yields accurate results for lines that are not strongly saturated. As an example of how the software fits the absorption lines we display in figure 2 the normalized PG 1634+706 spectrum together with the Gaussian fits made by the line finding and measurement software. Fits are displayed for all lines with $SL > 3.0$. Only lines with $SL > 4.5$ are included in the complete sample.

The G270H data obtained for PG 1634+706 constitute a special case since these data have signal-to-noise ratios comparable to the available flat-field data. This high sensitivity makes it difficult to determine if the numerous weak features that are detected are intrinsic to the spectrum or are residuals of partially-corrected flat-field features. For PG 1634+706, we imposed a minimum observed equivalent width of 0.09 Å for an absorption line to be included in the preliminary line list (described above) and a minimum equivalent width of 0.135 Å for inclusion in the complete sample.

The high line density in the $z \approx 1$ spectra caused a number of problems for the Paper II Gaussian fitting software, which was designed for and tested on spectra having the line density of the Ly-$\alpha$ forest at $z < 0.6$. The primary difficulty encountered was instability in the values of the line parameters that occurred when simultaneously fitting several (often more than five) lines. The software employed in this paper contains three line-fitting options (all used in this work) that were not present in the Paper II code: 1) the FWHMs of emission lines are fixed to be the FWHM of the SSF (most, if not all, of these weak emission lines are spurious; broad and/or strong emission lines are removed in the continuum fit), 2) the equivalent width of a given line is not allowed to change sign (e.g., an absorption line found in the SSF search cannot become an emission line in the Gaussian fit), and 3) a maximum allowable value of the FWHM can be set for weak components in blends containing more than three lines (15 Å in this paper). If a line violates either of the last two criteria in the fitting process, it is dropped from the line list.

If the Gaussian fitting software identified a feature with only a single component, we counted that feature as a single line, even though the FWHM of such lines are occasionally unphysically large. At higher resolution, we expect these broad features to break up into multiple components. Since blended, broad features are more likely to occur at the higher line densities found at the higher redshifts, we may be slightly undercounting the $z \sim 1$ systems relative to lower redshift systems. An accurate estimate of the size of this effect can be determined with the aid of the Monte Carlo simulations of our detection and identification sensitivities that we plan to undertake when all of the Key Project data have been analyzed.

The eye can sometimes be deceived concerning the significance of a particular absorption feature, which is a primary reason for selecting lines via a well-defined algorithm. The most subjective element in our analysis is the continuum fits in the vicinity of strong emission lines; these fits were done independently by at least three members of the Key Project in all cases (and merged into a consensus continuum by BTJ).

The new software was tested on more than 40,000 simulated spectra to determine the reliability of the measured line parameters and estimated errors. (The simulation software is described in Section 5 of Paper II). The simulated data have a SNR (40) and a resolution (4 pixels) similar to that of the G190H and G270H data discussed in this paper. We have expanded the Paper II simulation testing from the case of an isolated line (see Figure 12 in Paper II) to doublets and triplets whose separations are comparable to the widths of the lines. Unfortunately, the parameter space to be explored rapidly rises as the number of lines per blend increases. We have not exhausted all possibilities in the simulations performed to date; below we give a summary of some typical cases that are relevant to the data presented in this paper.

For isolated lines, the distribution of measured parameters was an excellent match to that expected from the analytic calculation. Spectra were created containing equal strength doublets (0.6 Å equivalent width) with a variety of separations. When the lines were separated by 1.5 times the line FWHM, more than 99% of the blends were split into two equal lines; this success rate dropped to 86% when the lines were only 1.25 FWHM apart. The program had a 99% success rate (i.e., split doublets into two lines) when measuring the properties of systems containing lines with 1.2 Å and 0.6 Å equivalent width separated by 1.75 FWHM. In a case of equal strength triplets (0.6 Å, 1.5 FWHM separation), 93% of the simulated spectra were correctly measured (split into three equal strength lines); if the equivalent widths of the lines were 0.6 Å, 1.2 Å, and 0.6 Å, the fraction of blends that were correctly deblended dropped to 81%.

The measured line centers and equivalent widths for the doublets and triplets were, on average, reasonably accurate (the offset between the mean measured and true values were much less than the $1\sigma$ errors). The estimated $1\sigma$ errors for the line centers were a good representation of the dispersions of the measurements, but the estimated $1\sigma$ errors for the equivalent widths was about 25% smaller than the measured dispersion. The discrepancy in the equivalent width error determinations was caused by a few lines with very large errors; the error estimate made by the program was a good description of the bulk of the simulations. The software overestimated the line widths and underestimated the $1\sigma$ line width errors. The bias towards large line widths is easily understood since this measurement is not symmetric; the measured FWHM can be up to 15 Å wide, but there is a limit to the smallest allowed width (the FWHM of the SSF). As is the case with the equivalent width measurements, most of the discrepancy is caused by a few lines with large errors.

Although we have a well-defined, automated procedure for finding and measuring absorption lines, the complexity of the spectra presented in this paper cause the output line list to be dependent on the search and fitting parameters. The SSF search is sensitive to both the $SL$ threshold and the number of iterations. In the Gaussian fitting, there are several regions where more than five lines are being simultaneously fit, and while the software modifications described above make the procedure much more robust, the fitting of these blends is dependent on the input parameters.

Initial trial line lists for three of the objects in this paper (Ton 153, PG 1352+011, and PKS 0122−00) were produced by the Paper II software. (This exercise revealed the inability of the original software to cope with regions of such high line density.) A comparison of these initial line lists with the output of the modified software (the "final line lists") gave a rough sense of the size of the systematic errors that can be expected in the line lists. For PKS 0122−00, the comparison is straightforward; both sets of line-measuring software were run on the same normalized spectrum. The line lists produced by the two software versions for the other two objects represent an extreme estimate of the systematic errors because modifications were made to the continuum fit between the two line-fitting sessions. Minor changes in the continuum fits can introduce large changes in the analysis of blends.

There were four types of significant differences between the line lists: 1) lines whose equivalent widths changed by more than twice the $1\sigma$ line width errors; 2) features that were one line in the Paper II software lists but were split into two in the final line lists (there were no examples of the reverse process); 3) weak lines in the original lists being dropped in the final line lists; and 4) weak lines in the final lists that were not in the original line lists. The number of lines in each category for the three objects were: Ton 153 (6, 1, 0, 6; out of 67 lines in the final list), PG 1352+011 (14, 3, 1, 10; 51 lines), and PKS 0122−00 (5, 0, 1, 5; 83 lines). While the modified software significantly improved the quality of the fits, it did not change greatly the output line lists. Only one of the line lists showed a change of more than 10% in the number of lines (PG 1352+011); a detailed comparison of the two line lists for this object revealed that over 95% of the differences could be attributed to the change in the continuum level.

The collection of lines having $SL > 4.5$ in a given spectrum will be referred to as the "complete sample." The minimum observed equivalent width as a function of wavelength that a line must have to be included in the complete sample is:

$$W_{\min}(\text{complete sample}; \lambda) = 4.5 \times \sigma_{\det}(\lambda) \quad , \tag{2}$$

i.e., $W > W_{\min}(\lambda)$. We have also constructed lists of lines with $3.0 < SL < 4.5$; these unpublished lists are referred to as "incomplete samples". We do not use the lines in the incomplete samples in our statistical analyses, although we occasionally remark on such lines in the notes for individual objects.

We have remeasured the redshifts of the LLSs in Ton 153 and PG 1352+011 (see Paper I) using a technique similar to, but slightly different from, that presented in Paper II. The modified software was also used to identify and measure two LLSs in PG 1634+706. A detailed description of the modifications to the Paper II software used for analyzing LLSs will be given in the next Key Project Survey catalog. For the purposes of the current paper, the most important impact of the software revision concerns the measurement of LLS redshifts. The redshifts of LLSs are determined here by assuming that 911.8 Å in the rest frame corresponds to the observed wavelength at which the continuum flux begins to rise (as one moves to longer wavelengths) from the depression due to the LLS. This prescription differs from the definition of a LLS redshift that was adopted in Paper II, for which 911.8 Å was assumed to correspond to the wavelength at which the flux was midway between the flux levels on the two sides of the LLS. The new procedure produces redshifts that are systematically about 0.02 lower than the redshifts given in Paper II. The current procedure yields redshifts that are in agreement with those identified from the narrow absorption lines to a characteristic accuracy of $\pm 0.003$, with no discernible systematic shift in the four cases reported in this paper (see discussion of LLSs in § 6.).

## 4. Identification of Absorption Lines

Our procedure for identifying absorption lines, which is embodied in an evolving software package called *ZSEARCH*, has been described in Paper I and in Bahcall et al. (1992). Our goal is to create a set of physically-sound identification algorithms that can be applied objectively and efficiently to analyze large numbers of simulated spectra that have the same characteristics as the observed data. In order to make the identifications in a way that can be analyzed statistically, the identification code must specify, among other things, the rules to be followed in deciding if a given set of identifications is consistent with atomic physics, the maximum allowed discrepancies between observed and calculated wavelengths, the range of candidate redshifts that are to be considered, and must properly take account of the measuring errors and detection sensitivities at each observed wavelength. In addition, one must specify a standard set of unredshifted absorption lines that are most likely to occur in gas anywhere between us and the quasar, as well as the order of preference if multiple identifications of the same observed line are possible with different standard lines. (For a recent, extensive list of short wavelength absorption lines see Verner, Barthel, & Tytler 1994).

We only outline here the salient aspects of the procedure we use and highlight improvements that have been made since Paper I was written. Many of these improvements are required by the order of magnitude higher absorption line density in portions of the high S/N spectra considered in this paper relative to the spectra considered in Paper I. We begin in this paper the process of evaluating the sensitivity and validity of our identification process by calculating the accidental probability of identifying pseudo-doublets (like C IV or O VI, see § 5.).

The standard ultraviolet absorption lines that we consider are the strongest allowed, one-electron, dipole transitions from ground or excited fine-structure states of cosmically abundant elements. Since the number of accidental coincidences increases with the number of standard lines considered, it is important to include only standard lines that are likely to have significant equivalent widths in our spectra. Therefore, we have used the standard line search list presented in Table 7 of Paper I with the following modifications that are based upon our experience with line strengths observed in the spectra that were analyzed in Paper I and with ground-based observations of large redshift objects. We have added to the standard search lines the singlet resonance line of N III $\lambda 989.80$ and have omitted the weak Si II $\lambda 1808.01$ Å and the weak Ar I $\lambda\lambda 1048.22, 1066.66$.

The maximum discrepancy allowed between an observed and a candidate line identification is either 1 Å or $3\sigma(\lambda)$, whichever is larger (where $\sigma(\lambda)$ is the wavelength measurement error in the line center). Each of the four spectra discussed in this paper was also analyzed with algorithms that permitted discrepancies as large as $3\sigma(\lambda)$, even if this discrepancy exceeded 1 Å. No physically-plausible identifications were found with this looser definition of the allowable mismatches. All redshifts from zero to 10,000 km s$^{-1}$ larger than the emission line redshift are considered.

*ZSEARCH* checks that each set of candidate absorption identifications satisfies atomic physics requirements regarding their relative strength if they are produced by the same ion. For example, *ZSEARCH* checks the relative strengths of the pair of lines of all the candidate doublets, the relative strengths of all of the Lyman lines, as well as the relative strengths of the C II, N II, Si II, S II, and Fe II lines. The requirements take account of the local sensitivity of our detection, as defined by Eq. (2). If a line is tentatively identified with the stronger component of a pair, then the minimum allowed equivalent width for the weaker component, $W_{\rm weak}({\rm min})$, is defined in terms of the observed equivalent width of the stronger member of the pair, $W_{\rm strong}$, the $1\sigma$ measurement error for the equivalent width, $\sigma_{\rm strong}$, and the $f$-values of the two transitions. The defining relation is

$$W_{\rm weak}({\rm min}) \;=\; \frac{f_{\rm weak}}{f_{\rm strong}} \times [W_{\rm strong} - 2\sigma_{\rm strong}] \quad . \tag{3}$$

If $W_{\rm weak}({\rm min})$ exceeds the detection threshold at the expected wavelength, but the predicted line is not found in the complete sample of absorption lines, the identification of the strong component of the pair is rejected. If, on the other hand, a line is tentatively identified with the weaker component of a pair, the minimum allowed equivalent width for the stronger component, $W_{\rm strong}({\rm min})$, is equal to the observed equivalent width of the weaker line minus a correction for the measuring errors. In this case, the defining relation is

$$W_{\rm strong}({\rm min}) \;=\; [W_{\rm weak} - \sigma_{\rm meas}] \quad , \tag{4}$$

where the measurement uncertainty is taken to be $\sigma_{\text{meas}} = (\sigma_{\text{strong}}^2 + \sigma_{\text{weak}}^2)^{1/2}$. In all cases, the identification is accepted if the predicted component of the pair lies outside the observed wavelength range. The excited fine-structure line of C II $\lambda$ 1335 Å is accepted only if the ground-state transitions at 1036 Å and 1334 Å both satisfy Eq. (4) with the excited-state line being regarded as the intrinsically weaker line.

The FWHMs of the lines are not used by *ZSEARCH* in any of the currently-implemented algorithms. We have delayed trying to incorporate this measured parameter into the identification process since the inferred value of the FWHM is sensitive to the specifics of the fitting routines. We cannot distinguish between a blend of several unresolved lines and a single, intrinsically broad line. We will calibrate in a future paper the effects of blending and of large intrinsic widths on our detection sensitivities to different populations by performing analyses of simulated spectra with the same line-fitting software and *ZSEARCH* software packages as are used to analyze the real spectra.

The identification of absorption lines consists of four phases. *ZSEARCH* first identifies the Galactic interstellar lines, which constitute a significant "background" within which lines of extragalactic origin must be recognized. Then every line in the spectrum that is not identified with a Galactic line is considered to be a candidate Ly-$\alpha$ line at the appropriate redshift, $z_{\text{cand}}$, provided $z_{\text{cand}}$ does not exceed the emission-line redshift by more than 10,000 km s$^{-1}$. The additional observed lines tentatively identified at $z_{\text{cand}}$ as metal lines or as members of the Lyman series are required to have relative equivalent widths consistent with their known $f$-values. Next, *ZSEARCH* performs independent searches for the strongest expected metallic doublets (C IV, N V, O VI, Mg II, Si IV, Al III, and Zn II) that are not found in the search based upon the assumption that each extragalactic line is Ly-$\alpha$. Finally, *ZSEARCH* examines individually all matches of standard lines with observed lines for candidate associated-absorption systems, i. e., for systems with redshifts within 3000 km s$^{-1}$ of the emission-line redshift.

The numerical redshifts given in this paper are calculated by iterating, for multiple-line systems, the redshift initially found by *ZSEARCH*. The software finds the redshift that minimizes $\chi^2$, calculated using the unweighted differences of the observed and the predicted wavelengths. The redshift with the minimum $\chi^2$ is adopted and a final set of identifications based upon this redshift is determined. In Paper I, no iteration was performed to determine the best-fit redshift.

We have implemented (for this paper) in *ZSEARCH* algorithmic tests according to which all singlet metal lines are accepted as identifications at a given redshift only if either Ly-$\alpha$ is not accessible in the wavelength region studied or, if Ly-$\alpha$ is accessible, then Ly-$\alpha$ is present at a strength at least equal to that of the metal line (with measuring uncertainties taken into account, as in the discussion above of pairs of lines from the same ion). In particular, we have required that Eq. (4) be satisfied with the weak line being the candidate metal line and Ly-$\alpha$ being the strong line. The singlet metal lines to which this algorithm applies are: C III $\lambda$977, N III $\lambda$989, N II $\lambda$1083, Fe III $\lambda$1122, S III $\lambda$1190, Si III $\lambda$1206, O I $\lambda$1302, and Al II $\lambda$1670. This algorithm was required in order to suppress the many accidental coincidences with these singlet lines for which no atomic physics self-consistency checks could be applied.

In order to further reduce the number of accidental identifications, we have also required that a number of other metal lines (with rest wavelengths in the general region of Ly-$\alpha$ ) only be accepted if either Ly-$\alpha$ is not accessible or, if accessible, Ly-$\alpha$ and the candidate metal line satisfy Eq. (4) with Ly-$\alpha$ being the strong line. The additional metal lines for which this test is applied in the current version of *ZSEARCH* are: Si II $\lambda\lambda$989, 1260, O VI $\lambda\lambda$1031, 1037, C II $\lambda\lambda$1036, 1334, N I $\lambda\lambda$1134, 1199, Fe II $\lambda$1144, N V $\lambda\lambda$1238,

1242, Si IV $\lambda\lambda$1393, 1402, C IV $\lambda\lambda$1548, 1550, and Al III $\lambda\lambda$1854, 1862.

So far, we have limited requirements on the relative strengths of metal lines from different ions since such constraints presume a knowledge of the ionization conditions or of the relative abundances. We have, however, used Eq. (4) to require that O I $\lambda$1302 be present if either N I $\lambda$1134 or N I $\lambda$1199 is to be accepted (with O I $\lambda$1302 being the strong line). We have also required that C IV $\lambda$1548 satisfy Eq. (4) in order for either of the presumably weaker Al III $\lambda$1854 or Al III $\lambda$1862 lines to be accepted. Both the O I versus N I requirement and the C IV versus Al III requirements were satisfied by all of the well-established absorption line systems we examined (in the present data and in the data presented in Paper I) and appear likely to be satisfied under rather general conditions.

No single-line identifications (one line per redshift) are allowed except for Ly-$\alpha$. Moreover, we have now implemented a software check that requires that at least two lines from the same ion be present if a candidate multiple-line redshift is to be accepted, unless the absorption redshift is within 3000 km s$^{-1}$ of the emission redshift. This rule eliminates a large number of accidental candidate redshifts in which one line each from a few different ions satisfies the other search algorithms.

*ZSEARCH* finds many cases where a single line is potentially identified at two or more redshifts with different standard lines. These cases of multiple potential identifications are resolved according to the following ordered set of priorities: 1) a Galactic interstellar line; 2) an extragalactic line in a metal-line system with an identified Ly-$\alpha$ line; and 3) an isolated Ly-$\alpha$ line. The reasons for these priorities are discussed below.

The probability that an extragalactic line falls very close to the position of a given Galactic interstellar line is relatively small for the spectra considered here, and experience has shown that absorption lines found at the positions of strong interstellar lines are nearly always Galactic in origin unless this conventional identification is inconsistent with the usual pattern of relative strengths of the commonly-occurring interstellar lines. Only the strongest Galactic lines are identifiable at the sensitivity levels reached by our spectra. For this paper, the interstellar identifications were examined individually and considered in the light of our experience in Papers I and III and in the light of the body of knowledge previously available from other satellite studies of ultraviolet absorption lines (Savage 1988). In the four spectra discussed in this paper, there were no unresolvable ambiguities regarding the validity of identifications with interstellar lines, although there were cases where a prominent absorption line in a multiple-line metal absorption system at a large redshift happened to fall near the rest wavelength of a potential interstellar line. Cases in which extragalactic absorption lines are identified at the observed wavelengths of strong interstellar lines are noted in the discussion of the spectra of individual objects in the text and reasons for the preferred identifications are given. For example, the O VI line ($W = 1.1$ Å) in the strong metal-line system at $z = 0.9531$ of Table 10 (referring to PKS 0122−00) falls on the ISM line at 2026 Å (which probably has only a minor contribution from the normally weak ISM Zn II or Mg I lines). Similarly, although the primary identification of the line observed at 2599.88 Å ($W = 1.9$ Å) is interstellar Fe II $\lambda$ 2600, an extragalactic line must also be present since the observed line at 2600 Å is many standard deviations stronger than the intrinsically stronger Fe II $\lambda$ 2382 ISM line.

The probability for accidentally identifying metal-line absorption systems that satisfy all of the self-consistency rules is small. Even if there are only three lines in a candidate redshift system, Ly-$\alpha$ plus a strong doublet, the probability of these three lines being a chance identification is typically less than 0.05 (see Table 4 and Table 5 in § 5.). In the most important cases considered here, several (and in some cases many) metal lines that fit a plausible ionization and abundance pattern and that satisfy all of the rules for the relative strengths of lines from the same ion are linked together in a single system. These multiple strong absorption systems are generally non-overlapping (as can be seen from an examination of Tables 6-16 of this paper), which is consistent with these metal-line systems not being produced by chance coincidences.

Absorption lines that lie shortward of the Ly-$\alpha$ emission line are presumed to be Ly-$\alpha$ or higher members of the Lyman series unless they are identified as Galactic interstellar lines or as metal lines in a multiple-line absorption system. An identification that is part of an absorption system with multiple lines is preferred algorithmically over an identification with only one other line.

For the four spectra discussed in this paper, the average line density (apart from strong Galactic ISM lines) shortward of Ly-$\alpha$ emission is 0.073 lines per Å, seven times larger than the average line density, 0.011 lines per Å, longward of Ly-$\alpha$ emission.

Blending of some spectral lines is inevitable at the resolution of these spectra ($R = 1300$). If one line appears to occur in more than one absorption redshift system with comparable plausibility, then we tabulate the dominant identification with the standard line and redshift but indicate by the notation *Bl* that the line is blended. We comment on the most important blends in the discussion sections referring to each object.

The reader interested in details of how the identification rules were applied in practice will find it useful to read the notes on individual spectra that are presented in (§ 6.).

## 5. Searches For Simulated Doublets

To estimate the probability of accidental redshift identifications that occur in spite of the imposition of the rules on the relative strengths of the lines (as discussed in § 4.), we have searched for pseudo-doublets in the observed spectra with the same algorithmic software that was used to make the line identifications described in the previous sections. To do this, we specified pseudo-pairs of absorption doublets with rest wavelengths, $\lambda_{1,\text{pseudo}}$ and $\lambda_{2,\text{pseudo}}$ that are produced from the real doublet atomic rest wavelengths, $\lambda_{1,\text{real}}$ and $\lambda_{2,\text{real}}$, with the aid of a random number generator. The relations used in creating the pseudo-doublets are:

$$\begin{aligned}
\lambda_{1,\text{pseudo}} &= \lambda_{1,\text{real}} \pm r \times \text{offset}, \\
\lambda_{2,\text{pseudo}} &= \lambda_{2,\text{real}} \pm r \times \text{offset}, \\
|\lambda_{1,\text{pseudo}} - \lambda_{2,\text{pseudo}}| &> \Delta\lambda,
\end{aligned} \quad (5)$$

where $\Delta\lambda$ is the minimum allowed separation.

Here $r$ is a random number between 0 and 1 and the offset parameter was taken to be 50 Å. The offset was either added or subtracted from the real wavelength depending upon whether a different random number, chosen also to be between 0 and 1, was larger or smaller than 0.5. No pseudo-doublet lines were allowed to have separations smaller than the separations of the doublet being simulated, i.e., the minimum allowed separation, $\Delta\lambda$, was 2.57 Å for the pseudo-C IV doublet and 5.68 Å for the pseudo-O VI doublet. We reversed the relative strengths ($f$-values) of the pseudo doublet pairs relative to the strength of the real pairs. Thus for C IV and O VI pseudo-doublets the longer wavelength line had an $f$-value that was twice as large as the $f$-value for the shorter wavelength line.

We have left in force all identifications of absorption systems found by *ZSEARCH* that have multiple-line identifications in the real spectrum, i.e., multiple-line metal systems and hydrogenic systems that contain at least Ly-$\alpha$ and Ly-$\beta$. The remaining lines that might be identified with real or pseudo-doublets are potential Ly-$\alpha$ lines (without Ly-$\beta$) and miscellaneous unidentified lines longward of the Ly-$\alpha$ emission line.

For each realization (i.e., Monte Carlo trial) of the atomic properties of a pseudo-doublet, we make use of a purged line list that is constructed as follows. From the actual line list of the observed spectra, we remove all the lines that belong to absorption systems that have secure multiple-line identifications, i. e., multiple-line metal systems which also contain Ly-$\alpha$ lines and Lyman systems that contain at least Ly-$\alpha$ and Ly-$\beta$ . The lines which remain are those that might be identified with real or pseudo-doublets; they are potential Ly-$\alpha$ lines (without Ly-$\beta$ ) and miscellaneous unidentified lines longward of the Ly-$\alpha$ emission line. In order to assess the probability of spurious C IV or O VI doublets being identified, we search this residual line list for occurrences of pseudo doublets using the same software, *ZSEARCH*, as is used in searching for real doublets in the complete observed line lists.

The procedure described here preserves all of the complexities in the observed spectra and in the identification algorithms while allowing us to estimate the probability of accidental identifications with pseudo-doublets.

The results are shown in Table 4 for pseudo C IV doublets and in Table 5 for pseudo O VI doublets. The first column lists the quasar whose spectrum was searched for pseudo-doublets and the second columns lists the number of potential lines that might be identified with real or (in the simulations) pseudo-doublets. The third column gives the average number of Ly-$\alpha$ plus pseudo-C IV (or Ly-$\alpha$ plus pseudo-O VI) identifications found per simulation. The average number was computed by dividing the total number of identified pseudo-systems found by the number (500) of simulations run. The fourth column gives the average number of pseudo-C IV (or pseudo-O VI) identifications per simulation that was found when we dropped the requirement that Ly-$\alpha$ be present when accessible. The last seven columns give the percentage of the total number of simulations in which 0 through 6 individual pseudo-C IV (or pseudo-O VI) identifications were found (without the requirement that Ly-$\alpha$ be present when accessible). For example, in the simulations run for Ton 153, about 51.6% had no identified pseudo-C IV identifications and 39.0% had exactly one identified pseudo-C IV identification.

There is a range in accidental probability of an order of magnitude depending upon which spectrum is being searched for pseudo-doublets. For pseudo C IV doublets, the average number per spectrum ranges from 0.05 for PKS 0122−00 to 2.35 for PG 1634+706. There is a remarkable 1% probability that more than 6 pseudo C IV doublets are found in the spectrum of PG 1634+706. All of the identified pseudo doublets lie within the Ly-$\alpha$ forest. For pseudo O VI doublets, the average number per spectrum varies from 0.002 for PKS 0122−00 to 0.046 for PG 1342+011. All of the identified O VI pseudo-doublets are associated with apparent Ly-$\alpha$ lines. The percentages given in Table 4 and Table 5 are based upon 500 different realizations of pairs of pseudo doublets that were created for each observed spectrum. We used a large number of simulations in order to oversample the parameter space. Therefore, the uncertainties in the estimated probabilities in Table 4 and Table 5 are larger than $1/(500)^{1/2}$.

The numbers of pseudo C IV doublets found in the observed spectra are sufficiently large that they must be taken into account in discussions of the identifications of the real absorption-line spectra. Note, however, that most of the pseudo C IV doublets that are identified are in regions of the observed spectra where the standard check by *ZSEARCH* on the strength of Ly-$\alpha$ is not applicable because the Ly-$\alpha$ absorption feature was not in an accessible part of the spectrum. Relatively few accidental O VI doublets are found in the observed spectra, principally because of the limited redshift range in which the O VI doublets are observable and also because Ly-$\alpha$ is nearly always accessible (and is required to be present) when O VI is detected.

## 6. Discussion of Individual Spectra

We discuss in this section the identifications of absorption lines in the individual spectra of Ton 153 (§ 6.1.), PKS 0122−00 (§ 6.2.), PG 1352+011(§ 6.3.), and PG 1634+706(§ 6.4.). The objects are arranged in order of increasing redshift.

### 6.1. Ton 153, $z_{\rm em} = 1.022$

The parameters of the lines found by our software are listed in Table 6. The individual columns in Table 6 contain: the measured line center; the $1\sigma$ error in the line center; the observed equivalent width; the $1\sigma$ error in the equivalent width; the significance level, $SL$ [see Eq. (1)]; the FWHM of the line; the identification, if any; the discrepancy, $\Delta\lambda$, in Å (observed minus predicted wavelength) between the proposed identification and the observed wavelength; and the absorption redshift. In total, 67 absorption lines were detected in the complete sample by the line selection software.

In the 800 Å shortward of Ly-$\alpha$ emission in this object, there are 57 absorption lines listed in Table 6 that are not Galactic features. There is only one unidentified line in the 800 Å longward of Ly-$\alpha$ emission. Therefore, nearly all of the lines shortward of Ly-$\alpha$ emission are expected to be associated with systems displaying Lyman absorption lines. We identify in Table 6 a total of 38 Lyman absorption systems, of which two systems ($z = 0.6606$ and $z = 1.0022$) have detected metal lines.

We have resolved a complex Ly-$\alpha$ absorption feature, with five Ly-$\alpha$ lines between 2029 Å and 2039 Å. There is an additional, weaker feature in the incomplete sample at 2026 Å, with $W = 0.18$ Å and $SL = 3.6$. The Ly-$\beta$ absorption corresponding to this complex is observed at 1713.56 Å, $z = 0.6706$, and is a broad feature with a FWHM of 3.3 Å. (It could also be blended with the stronger O VI line at $z = 0.6606$.) There are two Ly-$\alpha$ lines that correspond well to this Ly-$\beta$ redshift, at 2029.16 Å ($z = 0.6692$) and 2032.06 Å ($z = 0.6716$). Both of these lines are part of a larger, blended complex of five Ly-$\alpha$ lines in the range from 2029 Å to 2039 Å with redshifts extending from $z = 0.6692$ to $z = 0.6771$.

The broad Ly-$\beta$ complex very likely corresponds to the entire complex of five blended lines centered on 2034 Å and extending over $\approx 1400$ km s$^{-1}$. These results suggest that a Ly-$\alpha$ absorption system may exhibit velocity structure that contains individual components with relative velocities amounting to hundreds of km s$^{-1}$. It would be very useful to obtain GHRS spectra, with higher-resolution than the FOS spectra discussed here, of the features at 1713 Å and 2034 Å in order to clarify the nature of the suggested Ly-$\alpha$ complex (see also the discussion below of possible C IV features in the Ly-$\alpha$ forest).

No metal lines are detected from this complex of Ly-$\alpha$ absorption lines. For example, $4.5\sigma_{\text{det}}$ upper limits (in parentheses) on the equivalent widths of the following, often–strong metal lines are obtained: C II $\lambda$1335 line ($< 0.16$ Å), the C IV doublet lines ($< 0.37$ Å), N II $\lambda$1083 ($< 0.27$ Å), O I $\lambda$1302 ($< 0.16$ Å), Si II $\lambda$1260 ($< 0.16$ Å), and Si III $\lambda$1206 ($< 0.21$ Å). In the $z = 0.6716$ component, there is a possible identification of C II 1334 Å with the line observed at 2229 Å ($W = 0.22$ Å, identified as a Ly-$\alpha$ line in Table 6) and N II $\lambda$1083 with the line observed at 1811 Å ($W = 0.16$ Å, identified in Table 6 as Ly-$\beta$ at $z = 0.7667$). No other candidate metal lines appear at the redshift of $z = 0.6716$ (upper limits comparable to the ones given above for the $z = 0.6692$ system), which is why we think it likely that the two metal-line candidates that are found at $z = 0.6716$ are accidental coincidences.

There is a strong metal-line absorption system at $z = 0.6606$, which is separated by only 1600 km s$^{-1}$ from the broad Ly-$\alpha$ complex ($z = 0.6692$ to $z = 0.6771$). The estimated redshift of the LLS, $z_{\rm LLS} = 0.659$ (determined as described at the end of § 3. from the lower-resolution G160L spectrum discussed in Paper I), is in good agreement with the redshift of the multiple-line metal system. It is likely that the broad complex of five Ly-$\alpha$ lines, the strong metal-line system, and the LLS, are all one associated complex of very large scale, extending over 3000 km s$^{-1}$ (see discussion in § 8.).

We identify, in Table 7, 10 absorption lines at this redshift: the Ly-$\alpha$ and Ly-$\beta$ lines, two resonant lines of C II and one of C III, three strong lines of Si II and the resonant line of Si III. The stronger Fe II ($\lambda 1144$) line is blended at an observed wavelength of 1902.03 Å with the Ly-$\beta$ line of the $z = 0.8544$ system and the C II ($\lambda 1036$) line identified at 1720.57 Å is blended with the Ly-$\beta$ line at $z = 0.6671$. As is apparent from Table 7, the $z = 0.6606$ system is dominated by intermediate stages of ionization since O I, N I, C IV, N V, and Si IV are all absent. The Mg II doublet has been detected by Steidel & Sargent (1992) in ground-based spectra. It would be useful to search for the strong Fe II $\lambda\lambda 2586.65, 2600.17$ lines at the wavelengths (in air) of 4294.2 Å and 4316.6 Å. The upper limits on the equivalent widths of the strong features of O I, C IV, and O VI (blended with the Ly-$\beta$ line of the Ly-$\alpha$ complex discussed earlier) are, respectively: 0.15 Å, 0.36 Å, and 1.1 Å. A curve of growth analysis of this absorption could provide interesting information about the silicon to hydrogen abundance ratio, but would be more secure with higher-resolution (GHRS) studies.

The strong absorption system at $z = 0.6606$ is only about 2000 km s$^{-1}$ from the Ly-$\alpha$ complex centered at 2034 Å and—as indicated by the statistical discussion in § 8.—is probably part of the same extended gaseous complex.

The blended complex of lines centered on 2034 Å illustrates the difficulty of identifying metal-doublets, like C IV, in the Ly-$\alpha$ forest. The spacings and relative equivalent widths of this complex of lines are well fitted by two C IV doublets: one at $z = 0.3106$ ($\lambda_{obs} = 2029.16$, 2032.06 Å) and the other at $z = 0.3141$ ($\lambda_{obs} = 2034.49$, 2036.95 Å). The Ly-$\alpha$ lines that would correspond to these doublets are outside the region in which we have appreciable sensitivity. These doublets are rejected because of the information regarding the Ly-$\alpha$ complex and because the line at 2036 Å is part of a Ly-$\alpha$, Ly-$\beta$ pair at $z = 0.6754$. Another possible C IV doublet, with $z = 0.2891$ at 1995 Å and 1999 Å would—without other information—have been rejected because the 1995 Å line is potentially part of a Ly-$\alpha$ and Ly-$\beta$ pair at $z = 0.6444$. (The Ly-$\beta$ line at $z = 0.6444$ itself is blended with the Ly-$\gamma$ line at $z = 0.7339$ at the observed wavelength of 1686 Å.) However, Steidel & Sargent (1992) have detected in ground-based spectra the Mg II doublet corresponding to the $z = 0.2891$ system, which we therefore accept as real in Table 6. (In the statistical analysis of simulated spectra for which ground-based data were not available, lines like 1995 Å and 1999 Å would have been identified as Ly-$\alpha$ systems.) A possible O VI doublet pair at 2065 Å and 2077 Å, with accompanying Ly-$\alpha$ at 2434 Å, is accepted at the redshift, $z = 1.0022$, even though the C IV, N V, and S IV doublets are all absent. It would be interesting, but not conclusive, to search for the Fe II and Mg II lines at this redshift; they occur (in air) at: 5177.6 Å, 5204.6 Å (Fe II) and 5597.3 Å, 5611.7 Å (Mg II). In considering both the O VI identification and the C IV doublets, we have been guided by the Monte Carlo study of the frequency with which pseudo-doublets can be found in the spectra (see § 5.). For example, the chance that the O VI plus Ly-$\alpha$ identification at $z = 1.0022$ is accidental is only about 4% (see Table 5).

The interstellar line of Mg I $\lambda 2852.97$ is presumably present ($W = 0.47$ Å) at an observed wavelength of 2851.78 Å. The observed wavelength is poorly measured ($\sigma(\lambda) = 0.35$ Å) but is just far enough away (1.19 Å) from the standard wavelength not to be identified by the computer search. However, the weaker Mg I $\lambda 2026$ interstellar line is present ($W = 0.18$ Å) at the correct wavelength in the incomplete sample ($SL = 3.6$). The line at 3264 Å is probably not Mg II at $z = 0.1676$ since the (often strong) lines of Mn II, Fe II, Al III, C IV and Si II that could appear at this redshift are not observed. There is also no indication of the 3802 Å member of the Mg II doublet, although its implied strength, if 3264 Å is Mg II, would fall below that required to place it in our complete sample line list. The 3264 Å feature that is in question is at the edge of our usable spectrum; it would be useful to check on its reality with ground-based observations.

### 6.2. PKS 0122−00, $z_{\rm em} = 1.070$

The 51 detected absorption lines in the complete sample are listed in Table 8 together with their measured properties and identifications, all in the same format as used in Table 6 of the previous section. We identify in Table 8 a total of 20 Lyman absorption systems, of which three ($z = 0.9667$, $z = 0.9531$, and $z = 0.3989$) have detected metal lines. The two extensive metal-line systems at $z = 0.9667$ and $z = 0.9531$ are probably a manifestation of large scale structure since they are separated by only 2000 km s$^{-1}$. These two extraordinary redshift systems lie within 17 Å of each other, whereas we find only a total of 8 extensive metal-line systems in 1650 Å of accessible spectra. The *post facto* Poisson probability of finding two extensive metal-line systems this close together in our sample is less than 1 %. It would be interesting to obtain spectra of faint galaxies in this field in order to elucidate the dynamical properties of objects in what may be one of the earliest known examples of large scale structure.

The algorithmic software was not able to successfully deblend the broad complex of absorption that includes the lines at 2015 Å and 2017 Å without human intervention. The software was rerun in a restricted region centered on the complex of lines in question and a subjective decision was made as to when a satisfactory solution was reached.

The spectrum of PKS 0122−00 reveals two extensive high-ionization metal-line systems that are presented in Table 9 and Table 10, neither of which correspond to detectable LLSs. The upper limit to the optical depth for both the LLSs at $z = 0.9667$ and $z = 0.9531$ is $\tau_{\rm LLS} < 0.25$.

The metal-line system with $z_{\rm abs} = 0.9667$, shown in Table 9, contains 11 identified lines, including Ly-$\alpha$, Ly-$\beta$, Ly-$\gamma$ (blended with Ly-$\beta$ at $z = 0.8636$), Ly-$\delta$, and Ly-$\epsilon$ (blended with Ly-$\gamma$ at $z = 0.8949$), as well as the O VI doublet, the stronger lines from the C IV and Si IV doublets, and the C III and N III resonance lines. None of the absorption lines from low-ionization stages of C II, N I, N II, O I, Si II, S II, and Fe II are detected. The Ly-$\gamma$ line is blended with Ly-$\beta$ of $z = 0.8636$ at $\lambda 1991$ and the Ly-$\epsilon$ line is blended with Ly-$\gamma$ line of $z = 0.8949$ at $\lambda 1843$.

The metal-line system with $z_{\rm abs} = 0.9531$, shown in Table 10, also contains 11 high-ionization lines. This system contains Ly-$\alpha$, Ly-$\beta$, Ly-$\gamma$, Ly-$\delta$ (blended), and Ly-$\epsilon$ (blended), as well as the C IV and O VI doublets and the C III and the N III resonance lines. No lines are detected from low-ionization stages of C II, N I, N II, O I, Si II, S II, and Fe II. The Si III resonance line is not detected but the stronger component of the Si IV doublet is present just below the cutoff for the complete sample (with $SL = 4.44, W = 0.56$ Å) at 2722 Å. The Ly-$\delta$ line is blended with the Galactic interstellar line of Al III at 1854 Å and the Ly-$\epsilon$ line is blended with Ly-$\alpha$ of $z = 0.5069$ at 1831 Å.

The N V resonance lines are not detected in either of the two extensive metal-line systems, $z_{\rm abs} = 0.9531$ and $z_{\rm abs} = 0.9667$. Upper limits to the equivalent widths, 4.5 $\sigma_{\rm det}$, of the stronger N V lines are 0.65 Å and 0.69 Å. It would be useful to make ionization models of these systems to determine if they have an anomolously low abundance of nitrogen relative to carbon or oxygen. Lu & Savage (1993) have noted the weakness of N V relative to C IV and O VI absorption in the co-added spectra of large-redshift quasars observed from the ground.

There is a metal-line system at $z = 0.3989$ that contains Ly-$\alpha$ and a C IV doublet. The average number of pseudo-C IV doublets with accompanying Ly-$\alpha$ that are identified in the spectrum of PKS 0122−00 is only 0.002. This identification therefore appears secure even though the C IV doublet is present in the Ly-$\alpha$ forest.

It would be useful to search with ground-based observations for the strong Fe II and Mg II lines at the three metal-line redshifts discussed here: $z = 0.9667$, 0.9531, and 0.3989. The Fe II lines corresponding to these redshifts would appear, respectively, (in air) at: 5085.7 Å, 5112.3 Å; 5050.6 Å, 5077.0 Å; and 3617.4 Å, 3636.3 Å; and the Mg II lines would appear (in air) at: 5498.1 Å, 5512.2 Å; 5460.0 Å, 5474.1 Å; and 3911.7 Å, 3920.7 Å. One would not necessarily expect to find Mg II or Fe II with large equivalent widths because high ionization stages are predominant in the ultraviolet spectra of all three absorption systems.

The four strong lines that are detected in both the G190H and the G270H data do not show very good wavelength agreement; the differences in the line centers measured in the two gratings are: $+0.19$ Å, $-0.75$ Å, $+0.05$ Å, and $-0.91$ Å. The wavelength scales of the gratings were not shifted relative to each other because no interstellar lines are strong enough to be in the complete sample; the interstellar lines of Fe II at $\lambda\lambda 2600, 2344$ are both present just below the cut-off for a complete sample (at $SL = 4.3, 4.2$, respectively).

### 6.3. PG 1352+011, $z_{em} = 1.121$

The 83 detected absorption lines in the complete sample are listed in Table 11, together with their measured properties and identifications. The identifications include a total of 43 Lyman absorption systems, two of which ($z = 0.6677$ and $z = 0.5258$) have detected metal-lines.

There are two strong, extensive, metal-line absorption systems identified in the spectrum of PG 1352+011. One of these systems, $z = 0.6677$, shown in Table 12, contains 18 identified absorption lines from a broad range of ionization stages and corresponds to a LLS observed earlier at low resolution (Paper I). The second system, $z = 0.5258$, shown in Table 13, contains 13 identified lines, primarily from low-ionization ions. Both of the metal-line systems have neighboring clumps (in redshift space) of Ly-$\alpha$ systems that are discussed in § 8..

The $z = 0.6677$ system contains three Lyman lines (Ly-$\alpha$, Ly-$\beta$, Ly-$\gamma$), as well as a total of four other strong lines from the low-ionization ions of C II, N II, and Si II (see Table 12). The estimated redshift of the LLS, $z_{LLS} = 0.664$ (determined as described at the end of § 3. from the lower-resolution G160L spectrum discussed in Paper I) is in good agreement with the redshift, $z = 0.6677$, determined from the higher-resolution spectra of the 18-line system. It is likely that the Ly-$\beta$ line at 1710 Å contains contributions from two Ly-$\alpha$ lines: 2026.09 Å and 2027.60 Å.

Within 9 Å of the Ly-$\alpha$ line of the $z = 0.6677$ metal-line system, there are three additional Ly-$\alpha$ lines. It appears likely that the entire complex of four redshift systems from $z = 0.6608$ to $z = 0.6677$ are physically associated, corresponding to a 1200 km s$^{-1}$ complex. In addition, there are two lines at 2036 Å and 2037 Å that are identified in Table 11 as C II $\lambda$1334 and CII* $\lambda$1335 at $z = 0.5258$.

The intermediate stages of ionization are represented by the C III, N III, and Si III resonance lines. The doublets of C IV, N V, Si IV, and O VI, all of which represent high-ionization stages, are present. There are three stages of ionization identified for carbon (C II-C IV), for nitrogen (N II, N III, and N V), and for silicon (Si II-Si IV) among the observed lines in this system. The line at 1989 Å probably has a significant contribution from Si II $\lambda1193$ although there must be another important source of absorption at this wavelength since Si II $\lambda\lambda\lambda1190$, 1304, and 1526 are all missing. Consistent with the rules we have adopted, the line at 1989 Å is listed in the table as Ly-$\alpha$. With regard to blending, the observed line at 1650 Å is a blend of N III $\lambda989$ and Si II $\lambda989$; the N II line at 1807 Å is blended with Ly-$\beta$ at $z = 0.7628$; and the N V line observed at 2072 Å is blended with Ly-$\beta$ at $z = 1.0205$.

The $z = 0.5258$ system shown in Table 13 contains Ly-$\alpha$ and the strongest lines from the low-ionization ions of C II, O I, Si II, and Al II, but the S II triplet is missing. Lines from the higher-ionization ions of C IV, Si III, and Si IV are also identified. Blending is apparent for several absorption lines in this system. The Ly-$\alpha$ line, which is observed at 1855 Å, has an observed equivalent width of 4.0 Å, larger than the equivalent width of the Ly-$\alpha$ line in the $z = 0.6677$ LLS. This broad feature, with a $FWHM = 3.9$ Å, is not well fit by a single Gaussian and is probably a blend. The observed line at 1816 Å must contain contributions from S III and Si II $\lambda1190$ at $z = 0.5258$, but is stronger than would be expected if they were the sole contributors of the strength of the observed line. The line at 1816 Å has a FWHM of 3.3 Å and is not well fit by a single Gaussian, implying that it also is probably a blend. The equivalent width of this line is larger than the equivalent widths of any of the observed absorption lines of C II, C IV Al II, Si II, and Si IV in the $z = 0.5258$ system. In other well-studied absorption line systems, including those investigated in Paper I, the S III line is usually less strong than the lines of C II, C IV, Al II, Si II, and Si IV. Therefore, the table lists the primary identification of the line at 1816 Å as Ly-$\alpha$. The line identified as O I in Table 13 is probably a blend, since it has a FWHM of 7.3 Å, larger than Ly-$\alpha$ at this redshift or any other line in the complete sample of absorption lines for this quasar.

Absorption is apparently observed from the excited fine structure state of C II in the $z = 0.5258$ system at 2037 Å, but not from the excited fine structure states of Si II. This line could also be another Ly-$\alpha$ line in the clump near the $z = 0.6677$ metal-line system. The constraints on the local physical conditions that are implied by the identification with the excited fine structure state of C II depend upon whether the excitation is predominantly via hydrogen atoms or predominantly via electron collisions (see Bahcall & Wolf 1968; Spitzer & Fitzpatrick 1993). If the C II identification and not the Ly-$\alpha$ identification is correct for the line at 2037 Å, then more accurate measurements of the relative strengths of the ground and excited fine structure states might permit estimates of the cooling rate in the gas which could be compared with estimates for the Milky Way halo gas in the direction of 3C 273 (see Savage et al. 1993a).

There are an additional four Ly-$\alpha$ lines within 12 Å of the Ly-$\alpha$ line of the $z = 0.5258$ absorption system. It seems likely that the absorbing gas between $z = 0.5161$ to $z = 0.5258$, containing a total of five Ly-$\alpha$ lines in the complete sample, constitutes a physically-related complex extending over 1900 km s$^{-1}$(see discussion in § 8.). The broad Ly-$\alpha$ line of the $z = 0.5258$ system may also itself show more than one component at higher resolution.

Given the detections of low ionization stages in the ultraviolet spectra, it seems probable that the strong Fe II and Mg II absorption systems would be detectable in ground-based observations of the $z = 0.6677$ and the $z = 0.5258$ absorption systems. The Fe II lines will occur (in air) at: 4312.5 Å, 4335.1 Å; and 3945.6 Å, 3966.2 Å; and the Mg II

lines will occur (in air) at: 4662.2 Å, 4674.1 Å; and 4265.5 Å, 4276.4 Å.

There are systematic uncertainties in identifying lines in the crowded region below Ly-$\alpha$ emission, particularly in the presence of broad, obviously blended features. *ZSEARCH* finds two candidate C IV doublets among the six blended lines in the complex between 1841 Å and 1855 Å that are not obviously identified in other systems. These two candidate doublets (at $z = 0.1925$ and $z = 0.1942$) involve the three observed lines at 1846 Å, 1848 Å, and 1855 Å. Since they share the 1848 Å line, only one of the candidate doublets can be real. The $z = 0.1942$ candidate system has a doublet ratio of equivalent widths of 0.38/0.62, which is to be compared with the optically thin ratio of 2 : 1 and therefore this candidate system is less likely to be real. The Monte Carlo simulations discussed in (§ 5.) suggest that there is approximately a 47% chance that the other candidate C IV doublet, $z = 0.1925$, is real. We have identified in Table 11 all five of the observed lines in question as Ly-$\alpha$ lines, but in estimating the Ly-$\alpha$ number density we have assumed that there is a 47% chance that the lines observed at 1846 Å and 1848 Å are a C IV doublet rather than a pair of Ly-$\alpha$ lines. The two lines, $\lambda\lambda 1846, 1848$ are labeled with a "p" in Table 11 in order to indicate that they were assigned in the Monte Carlo calculations a specific probability, less than unity, of being a Ly-$\alpha$ line (see also the discussion in § 7.).

The lines produced in the Galactic interstellar medium are relatively weak along this line of sight ($b = +60°$). Only the two strongest lines of Fe II, the Mg II doublet, the Al II resonance line, and the stronger Mg I line are definitely identified. There are observed lines at 2026 Å and 2062 Å, which coincide in wavelength with resonance lines of Zn II. However, the observed line at 2026 Å is one of the strongest absorption lines in the spectrum of PG 1352+011, stronger than any other Galactic interstellar line identified along this line of sight. The observed line at 2062 Å has a larger equivalent width than either of the two strongest Fe II interstellar medium lines. For the 13 lines of sight discussed in Paper I, the Fe II interstellar absorption lines are stronger than the Zn II lines. Therefore, the Zn II interstellar medium identifications were rejected in favor of Ly-$\alpha$ lines.

The line at 2606 Å could be a Galactic interstellar Mn II line; this identification is consistent with the rules of *ZSEARCH*. However, the fact that the interstellar Mg II and Fe II lines are not strong along the PG 1352+011 line of sight argues against this possibility as the full interpretation of the observed line strength ($W = 0.26$ Å). For 10 of the 13 lines of sight studied in Paper I, no Mn II lines are detected although the interstellar Fe II lines are in all cases stronger than in PG 1352+011 spectrum. In the other sightlines of Paper I in which the interstellar lines are very strong and Mn II is detected, the equivalent width of the line at $\lambda 2606$ Å is less than the 0.26 Å equivalent width measured for PG 1352+011. A close examination of the spectrum in the vicinity of 2606 Å shows that the feature in question may be simply an artifact of a slightly imprecise continuum placement. Therefore, this line is listed in Table 12 as Mn II with a question mark.

## 6.4. PG 1634+706, $z_{em} = 1.334$

There are 90 absorption lines in the complete sample for this spectrum, 71 of which are in the 550 Å shortward of Ly-$\alpha$ emission and only 19 of which are in the 500 Å longward of Ly-$\alpha$ emission. Because the signal-to-noise ratio in these data is comparable to the sensitivity obtained in the calibration flat-fields, we have imposed a minimum equivalent width requirement of 0.135 Å for inclusion in the complete sample (see discussion in § 3.).

The complete sample of absorption lines are shown in Table 14 together with the identifications made with *ZSEARCH*. We identify in Table 14 a total of 47 Ly-$\alpha$ systems of which three systems ($z = 0.9060$, $z = 0.9908$, and $z = 1.0417$) have detected metal lines. We also measure (as described at the end of § 3.) the redshifts of two LLSs that are visible in the HST-archived G190H spectrum of PG 1634+706 that was taken by Impey, Malkan, & collaborators (1994). These two LLSs correspond to multiple-line metal absorption systems, $z = 1.0417$ and $z = 0.9908$, that are discussed below.

The strongest absorption line in the spectrum, Ly-$\alpha$ at $z = 1.0417$, corresponds to a LLS that is apparent in the IUE spectrum of this object (see Bechtold et al. 1984). The estimated redshift of the LLS as determined by our measurements, $z_{LLS} = 1.045$, is in good agreement with the multiple-line redshift determined from the G270H spectrum shown in this paper and the measurement of Bechtold et al. (1984). The identifications for this LLS are listed in Table 15. Altogether, the metal-line absorption system corresponding to the LLS contains eight lines in the complete sample, including the C IV and Si IV doublets, the strongest line of C II, the resonance lines of C III and S III, as well as Ly-$\alpha$. In addition, the strongest line of Si II $\lambda 1260$, and the strongest accessible line of Fe II $\lambda 1144$, are both present in the incomplete sample with $SL$ ratios of 4.1 ($W = 0.11$ Å) and 4.3 ($W = 0.14$ Å), respectively. The metal lines in the complete sample have equivalent widths ranging from 0.2 Å to 0.9 Å. The ground-based observations show that the Mg II doublet lines are also present with equivalent widths of 0.14 Å and 0.10 Å (see Bechtold et al. 1984).

The metal system at $z = 0.9908$, shown in Table 16, exhibits 11 lines in the complete sample, including Ly-$\alpha$, the C IV and Si IV doublets, two lines of Si II, and separate lines of C II, Fe II, Si III, and S III. The Mg II doublet corresponding to the $z = 0.9908$ absorption system was detected in ground-based spectra by Bechtold et al. (1984) and by Steidel & Sargent (1992). The estimated redshift of the Lyman-limit system, $z_{LLS} = 0.993$, is in good agreement with the metal-line redshift.

There is a cluster of three additional Ly-$\alpha$ lines within 15 Å of the Ly-$\alpha$ line of the $z = 0.9908$ system. These four systems, from $z = 0.9785$ to $z = 0.9908$ form a gaseous complex spanning 1900 km s$^{-1}$. The statistical significance of this cluster is not as high as the clusters previously identified and remarked upon in the spectra of Ton 153 and PG 1352+011.

There is also a relatively high-ionization metal-line absorption system at $z = 0.9060$ that contains six lines: Ly-$\alpha$, the C IV and N V doublets, and Si III (see Table 14). Lines from the lower-ionization ions of Si II, Al II, Fe II, N I, and O I are all absent, but the Si IV lines are also not observed. The C II $\lambda$1335 line is present in the incomplete sample at a $SL$ of 4.2 ($W = 0.12$ Å).

It would be useful to search with ground-based observations for the strong Fe II and Mg II lines at the three metal-line redshifts: $z = 1.0417, 0.9908$, and $0.9060$. The Fe II lines corresponding to these redshifts would appear (in air), respectively, at: 5279.7 Å, 5307.3 Å; 5148.1 Å, 5175.0 Å; and 4928.2 Å, 4954.5 Å; and the Mg II lines would appear at 5707.7 Å, 5722.4 Å; 5565.4 Å, 5579.7 Å; and 5328.4 Å, 5342.0 Å. One would expect the Fe II and Mg II lines to be detectable in the $z = 1.0417$ and $z = 0.9908$ systems since the ultraviolet spectra show the presence of other low-ionization ions. The absence of lines from low-ionization ions in the ultraviolet makes the detection of Fe II and Mg II less likely for the $z = 0.9060$ system.

There is an interesting candidate redshift system at $z = 1.3413$, approximately $10^3$ km s$^{-1}$ larger than the emission-line redshift, that contains eight candidate lines. This candidate redshift is probably not real. The candidate identifications are: the O VI doublet and singlet lines from Fe II, Ly-$\alpha$, O I, Si II, Si IV, and S III. The weaker component of the O VI doublet is otherwise identified as S III in the LLS at $z = 1.0417$ (see Table 15). Thus there is no ion from which two or more lines are detected in this candidate redshift. In addition, Ly-$\alpha$ at $z = 1.3413$ is identified as Si IV in the LLS and Si II $\lambda$1260 is identified as C IV at $z = 0.9060$ (see Table 14). Finally, six of the eight candidate identifications at this redshift have wavelength discrepancies larger than $3\sigma(\lambda)$. We have therefore rejected this candidate system in favor of the listed identifications.

There are four candidate C IV doublets in the Ly-$\alpha$ forest of this spectrum with redshifts equal to 0.5582, 0.6540, 0.6790, and 0.7796. These doublets correspond to the following observed lines: 2412 Å and 2416 Å, 2560 Å and 2564 Å, 2599 Å (which is probably predominantly interstellar Fe II) and 2603 Å, 2755 Å, and 2759 Å. According to the simulations described in (§ 5.), there is an 18% probability that all four of these candidate doublets are the result of accidental coincidences and are not real. There is also a 7% probability that all four candidate doublets are real. For simplicity of presentation, we list all eight of the lines in question as Ly-$\alpha$ in Table 14, but in computing the number density of Ly-$\alpha$ absorption–line systems we attach a probability of 41% (derived from the simulations described in § 5.) that each of the 8 lines is a Ly-$\alpha$ line. The probable number of C IV lines is then $4 \times (1 - 0.41) = 2.36$, in agreement with the last row of Table 4. The eight lines in question are all labeled as "p" in Table 14 in order to indicate that they were assigned in the Monte Carlo calculations a specific probability, less than unity, of being a Ly-$\alpha$ line (see also the discussion in § 7.).

It is also possible that the lines at 3159 Å and 3164 Å, which are part of the same blend as the $z = 1.0417$ C IV doublet identified at 3161 Å and 3166 Å, are also a C IV doublet. However, this possibility is rejected by *ZSEARCH* since the corresponding Ly-$\alpha$ line is not detected. A further complication is that the average separation of the four blended lines mentioned above is 2.5 Å, which is essentially equal to the spectroscopic resolution, 2.4 Å. Higher resolution spectra are required to determine how many real lines belong to this complex.

In addition to the lines at 3159 Å and 3164 Å just just discussed, there are six unidentified lines in the 400 Å above the Ly-$\alpha$ forest that is covered in the G270H spectrum. These long-wavelength unidentified lines are relatively weak with an average equivalent width of only 0.28 Å. It is possible that some or all of these lines are flat-field features.

There are 42 lines in the 600 Å below the inception of the Ly-$\alpha$ forest that have no obvious identification other than Ly-$\alpha$ lines. The otherwise-unidentified lines in the Ly-$\alpha$ forest are relatively strong with an average equivalent width of 0.54 Å, almost twice the average strength of the unidentified lines outside the Ly-$\alpha$ forest. Making the conservative assumption that the average number of unidentified lines per Å that are not Ly-$\alpha$ lines is the same within and above the Ly-$\alpha$ forest, we estimate that 12 of the 42 Ly-$\alpha$ lines in Table 14 are misidentified. Making the further assumption that on average two of the candidate C IV doublets in the Ly-$\alpha$ forest are real and are not among the previously discussed 12 probable misidentifications of Ly-$\alpha$ lines, we estimate–combining the uncertainties quadratically–that as many as 13 of the 42, or 31%, of the Ly-$\alpha$ lines listed in the table could be misidentified. Since this estimate ignores the difference in average equivalent width between the usually stronger candidate Ly-$\alpha$ lines and the weaker longer-wavelength metals, the derived number (31%) significantly overestimates the probability of misidentifications.

The uncertainties in the Ly-$\alpha$ number density at the relatively larger equivalent widths that are usually reported in the literature are not very large since only one of the eight unidentified large wavelength lines has a rest equivalent width above 0.32 Å (at $z_{\text{abs}} \geq 0.15$). This translates to 1.5 misidentifications out of 15 strong Ly-$\alpha$ lines. Similarly, only three of the eight candidate C IV doublet lines in the forest are strong enough to possibly correspond to Ly-$\alpha$ lines with rest equivalent widths in excess of 0.32 Å. Assuming, in accordance with the simulations discussed in (§ 5.), that half of the doublet lines may be real, we estimate a doublet contamination of 1.5 lines in the Ly-$\alpha$ sample. Altogether, the probable number of misidentifications is about 2.2 lines, or 15%, of the candidate Ly-$\alpha$ lines with rest equivalent widths above 0.32 Å.

This line-of-sight ($b = +37°$ ; $\ell = 103°$) exhibits the strongest set of Galactic interstellar absorption lines that we have studied in the Key Project so far. All seven of the accessible Fe II resonance lines listed in the table of standard lines (Table 7 of Paper I) are present and strong. Although the primary identification of the line detected at 2599.88 Å is Fe II $\lambda 2600$, the observed line must be blended with one other line since the strength of 2599.88 Å (1.9 Å equivalent width) exceeds that of the intrinsically stronger Fe II $\lambda$ 2382 line (1.3 Å equivalent width) by many standard deviations. We have indicated in Table 14 that the line in question is probably an extragalactic Ly-$\alpha$ feature plus interstellar Fe II. In addition to the practically ubiquitous lines of interstellar Mg II and Mg I, all three resonance lines of Mn II are present.

## 7. Ly-$\alpha$ absorption lines

The addition of the redshift systems in the four quasars reported on above allow us to extend the investigation of the evolution of the Ly-$\alpha$ absorption systems up to a redshift of $\sim 1.3$. In Paper I, we assumed a distribution of the Ly-$\alpha$ absorption systems in rest equivalent width-redshift space of the form

$$\frac{\partial^2 N}{\partial z \partial W} = \left.\frac{dN}{dz}\right|_0 (W^*)^{-1} (1+z)^\gamma \exp(-W/W^*) \quad . \tag{6}$$

As discussed recently by Press & Rybicki (1993), an approximate exponential distribution of rest equivalent widths is a natural consequence of the convolution of the distribution of HI column densities and Doppler parameters actually observed at high resolution (see also Schneider, Schmidt, & Gunn 1989). This approximation ceases to be valid when the lines begin to be unsaturated, which, for typical Doppler parameters, occurs for rest equivalent widths in the range 0.16–0.12 Å. A significant number of the Ly-$\alpha$ lines in our new sample have rest equivalent widths less than 0.16 Å and there is evidence in our full sample that a single exponential distribution is not an adequate representation of the complete range of rest equivalent widths.

A more conservative procedure than using Eq. (6) is to adopt a minimum rest equivalent width for the entire sample, $W_{\min}$, and to restrict our analysis to the spectral regions in each quasar for which the 4.5$\sigma$ detection limit is less than $W_{\min}$. Only lines in regions with $W > W_{\min}$ are included in this conservative sample. We refer to such a sample as a "uniform" detection-limit sample. The line densities derived by this procedure are independent of any assumption about the equivalent width distribution, but are derived from a sample of reduced size.

Considering all of the 17 Key Project observations to date, including the 13 smaller redshift objects analyzed in Paper I and the 4 quasars discussed in the present paper, we find that the maximum number of lines for such a uniform sample occurs for $W_{\min} \sim 0.24$ Å. Restricting the sample further to Ly-$\alpha$ absorption lines having velocities more than 3000 km s$^{-1}$ different from the emission line redshift and differing by more than 300 km s$^{-1}$ from the absorption systems having heavy elements, there are 109 lines in the uniform sample. For $W_{\min} \ll 0.24$ Å, there are too few spectral regions having the required high SL ratio, while for $W_{\min} \gg 0.24$ Å, Ly-$\alpha$ lines are intrinsically rare.

For 10 of the lines marked by "p" in Table 11 and Table 14, we use probabilities that the lines in question are actually Ly-$\alpha$ lines. The values of these probabilities were calculated from the discussion in § 5. and § 6., see especially Table 4. The two lines marked "p" in Table 11 were assigned a probability of 53% that they were Ly-$\alpha$ lines and the eight lines marked "p" in Table 14 were each assigned a probability of 41% that they were Ly-$\alpha$ lines. The considerations that led to these probabilities are described in the notes referring to the particular spectrum in question in § 6.

We have used a slight modification of the maximum likelihood formalism described by Murdock at al. (1986). These modifications were described briefly in § 6.2 of Paper I. With this formalism, we find, for the local line density at zero redshift for lines with equivalent widths greater than 0.24 Å, $(dN/dz)_0 = 24.3 \pm 6.6$ Ly-$\alpha$ lines per unit redshift, with $\gamma = 0.58 \pm 0.50$. The uncertainties quoted here are 1-$\sigma$ values. They are derived in the approximation in which terms higher than second order in the expansion of the log of the probability (as a function of the free parameters) about its maximum value are neglected. This distribution is shown as the solid line of Figure 3. We have, after the fact, confirmed that for this sample an exponential distribution for the rest equivalent widths with $W^* = (0.21 \pm 0.02)$ Å is a good representation of the equivalent width distribution. The value of $W^*$ found here is virtually identical to the value found in Paper I and is somewhat lower than found in most analyses of high-redshift Ly-$\alpha$ clouds. The measured values of $W^*$ found at large redshifts may be biased toward larger values because of line blending (see Lu 1991).

Including the set of all lines in our sample with rest equivalent widths greater than 0.24 Å, regardless of the detection limit at the wavelength in question, increases the sample size by about 50%. The maximum-likelihood estimation procedure described in Paper I yields for this variable-detection limit sample rather similar results: $(dN/dz)_0 = 25.5 \pm 5.3$ Ly-$\alpha$ absorption lines per unit redshift, with $\gamma = 0.42 \pm 0.42$. This distribution is shown as the dashed line in Figure 3. Note that the line densities given above refer to all lines with rest equivalent width exceeding 0.24 Å. To obtain the number in excess of the commonly-used fiducial value of 0.32 Å, the line densities quoted above should be reduced by a factor of $\exp[-(0.32 - 0.24)/0.21] = 0.68$. With this correction factor to refer to lines stronger than 0.32 Å, the above-quoted line densities become respectively, $(dN/dz)_0 = 16.5$ and 17.3 Ly-$\alpha$ lines per unit redshift, in excellent agreement with the results, for only small redshifts, of Paper I [$(dN/dz)_0 = 17.7 \pm 4.3$ Ly-$\alpha$ lines per unit redshift and $\gamma = 0.60 \pm 0.62$] and Bahcall et al. (1993b) [$(dN/dz)_0 = 15 \pm 4$ Ly-$\alpha$ lines per unit redshift, for an assumed value of $\gamma = 0.75$].

We also show the line densities, $dN/dz$, for the uniform detection-limit sample, at the midpoints of each of 4 redshift bins over the interval $z = 0.0$ to 1.3. (For these estimates, we assumed a value of $\gamma$ that is found for the entire redshift region, 0.58, but the results are insensitive to this assumption.) The results are also plotted in Figure 3, where the vertical error bars represent only the Poisson errors associated with the number of Ly-$\alpha$ lines in each bin.

The rather gentle rate of evolution for the low redshift Ly-$\alpha$ lines which we find is at variance with most analyses of ground-based data referring to much higher redshifts. In Paper I we compared our results with the sample of Lu et al. (1991), which was based upon intermediate resolution ($\sim 150$ km/sec) data. A more recent analysis of a larger and somewhat higher resolution data set $\sim 50 - 75$ km/sec) has recently been provided by Bechtold (1995; and references therein) who finds a value of $\gamma = 1.71 \pm 0.23$ over the redshift range $1.6 \leq z \leq 4.0$ but notes that this simple power law fit does not fit the data very well.

There is some disagreement over the effect of blending when using intermediate resolution data to estimate $\gamma$ (cf. Bechtold 1994 and references therein and Trevese, Giallongo, & Camurani 1992.) In particular, Trevese et al. find that differential evolution (i.e. a $\gamma$ which depends upon line strength) is also strongly affected by line blending. A corollary of this is that the value of $W^*$ will increase (Lu 1991; Trevese 1992) over that found with higher resolution data. In particular our value of $W^* = 0.21$ is almost exactly the same as that suggested by Giallongo et al. (1991) for high redshifts, based upon high resolution data.

Even some recent echelle resolution data give somewhat ambiguous results. A recent summary is given by Williger et al. 1994 (section 4.2). Over the redshift range from 1.86 to 4.3, the best fit for all lines with log $(N_{\rm NI}) > 13.3$ gives $\gamma = 2.0 \pm 0.3$, however these authors note that *even at their resolution of* $\sim 10$ *km/sec*, some of these lines will be lost to blends at the higher redshifts. Confining their attention to lines with log $(N_{\rm HI}) > 14.5$ they obtain $\gamma = 4.6 \pm 0.7$ over the same redshift range. However, they suggest that $\gamma$ may steepen as the redshift increases over the range available to ground-based observations.

In an attempt to avoid issues of line blending, Press et al. (1993) recently analyzed high redshift, low resolution data with a procedure which avoided counting individual lines. It therefore refers to the cumulative effect of lines of all strengths. Because low resolution data was used, the estimate of the continuum shortward of the Ly-$\alpha$ emission lines must be based upon extrapolation of the continuum from longer wavelengths. Over the redshift range 2.5–4.3, Press et al. obtain $\gamma = 2.46 \pm 0.37$. More recently, Storrie-Lombardi (1994) applied a similar technique to a much larger data set (including some objects observed at intermediate resolution and high signal to noise) and found $\gamma = 2.56 \pm 0.11$ over the redshift range $2.5 \leq z \leq 4.9$.

Comparing our results for $REW > 0.24$ ($\gamma = 0.42 \pm 0.42$ and $\gamma = 0.58 \pm 0.5$ for the variable and uniform detection limit samples, respectively) with the foregoing high redshift samples, one finds that the differences between the FOS and the ground-based data sets in the $\gamma$'s are significant at the $\sim$ 2–4.5$\sigma$ level, depending upon the particular high redshift result selected for the comparison.

Thus, there is moderately strong, though not yet compelling, evidence that the overall rate of evolution of the Ly-$\alpha$ forest line density changes between the range of redshifts studied in Paper I and this paper (0–1.3) and that to which the ground based data discussed above refers ($\sim$ 1.6–4.9). What is not yet clear is whether this change is gradual or changes rather abruptly at one or more epochs, and whether the rate of evolution depends on line strength.

## 8. Clumping of Ly-$\alpha$ Lines Near Extensive Metal-Line Systems

In this section, we describe and analyze the clumping of Ly-$\alpha$ absorption systems near extensive metal-line absorption systems that are found at redshifts between 0.5 and 1.0. We interpret this clumping as providing evidence that some Ly-$\alpha$ absorption lines are associated with galaxies since it is generally believed that many or most of the metal-line systems are associated with galaxies. Previous evidence that some fraction of the small-redshift Ly-$\alpha$ absorption lines are associated with galaxies and/or clusters of galaxies has been inferred from observations of H1821 + 643 (Bahcall et al. 1992), PKS0405-123 (Spinrad et al. 1993) and from observations of 3C 273 (Bahcall et al. 1991a,b; Morris et al. 1991,1993), whose line of sight passes close to the Virgo cluster. Recently, Lanzetta et al. (1995) have also presented evidence for the association of a significant fraction of the low redshift Ly-$\alpha$ absorption lines with galaxies. Barcons and Webb (1990) did not find significant clustering of Ly-$\alpha$ lines around large redshift metal-line systems observed at 1 Å resolution.

Table 17 summarizes some of the most important facts concerning the extensive metal-line systems and the clumping of Ly-$\alpha$ lines. The second column of the table lists the metal-line redshifts of all eight extensive metal-line systems that we have identified in this paper; the third column gives the measured redshifts of detected LLSs, determined by the procedure described in § 3. The fourth column lists the number of metal lines in each system that are identified in the complete absorption-line sample and the fifth column describes the characteristic ionization level of the absorbing ions. The last three columns, columns six through eight, summarize the number of clustered Ly-$\alpha$ lines (including the Ly-$\alpha$ line from the Lyman-limit system) that are found close to the metal-line redshift, as well as the velocity dispersion and the apparent velocity span, of the entire complex of Ly-$\alpha$ lines and extensive metal-line system. The results given in columns four through eight are based upon the detailed analyses, described in § 6., of the spectra of the four quasars listed in the first column of Table 17. (It should be noted that line identifications were completed and a final manuscript of the paper was being prepared before the clumping of Ly-$\alpha$ lines was noticed.) In computing the velocity spans and the velocity dispersions listed in Table 17, we have included the Ly-$\alpha$ line of the neighboring metal-line system since, as argued below, we believe the metal-systems are physically associated with the Ly-$\alpha$ clumps.

The average separation of Ly-$\alpha$ lines is 22 Å, 0.0453 Ly-$\alpha$ lines per Å, for the complete sample of measured lines found in the spectra of the four intermediate redshift quasars studied in this paper. There are, by contrast, remarkable clumps of Ly-$\alpha$ lines shown in Table 17: 5 lines in a 10 Å span in the spectrum of Ton 153; 4 lines in 9 Å, and 5 lines in 12 Å, in the spectrum of PG 1352+011.

The spectrum of Ton 153 in particular shows five clustered Ly-$\alpha$ lines in the complete sample within a span of less than 10 Å from 2029.16 Å to 2038.76 Å (see discussion in § 6.1.; a sixth line is present at 2026 Å in the incomplete sample). The five blended (but resolved in our data) Ly-$\alpha$ lines extend over a velocity range of 1400 km s$^{-1}$ ($z$ = 0.6692 to $z$ = 0.6771) in the 10 Å centered on the observed wavelength of 2034 Å. No associated metal lines are detected from this clump of Ly-$\alpha$ lines (see § 6.1. for upper limits).

The "clumps" of Ly-$\alpha$ lines recognized in the spectra of Ton 153 and of PG 1352+011 are immediately apparent when displayed as shown in Figure 4 (see discussion below). They were 'discovered' *post facto* without the benefit of a rigorous definition of what constitutes a clump of lines. It is necessary to take this *post facto* nature of the identification of clumps into account when evaluating their statistical significance. Press and Schechter (1974) discussed a formalism that is appropriate for the statistical analysis of such a situation; they give expressions for the probability of a clump of lines with $\geq k$ members having some inferred local overdensity of lines being found in a sample of $N$ lines. The Press-Schechter formalism yields a probability of about 4 % for the chance occurence of the clump of 5 Ly-$\alpha$ lines in the spectrum of Ton 153.

The extensive metal-line system appears at 2018.83 Å, within 16 Å of the center of the Ly-$\alpha$ clump. At $z$ = 0.6606, the metal-line system is within 1600 km s$^{-1}$ of the shortest wavelength member of the Ly-$\alpha$ concentration. Since there is only one extensive metal-line system in more than 800 Å of accessible spectrum, it seems unlikely that the two most remarkable features in the spectrum, the clump of five Ly-$\alpha$ lines and the extensive metal-line system, should be accidentally so close to each other. The probability that the Ly-$\alpha$ line of the extensive metal-line system should fall by random chance as close as observed, within 16 Å, of the wavelength center ( 2034 Å) of the clump of the Ly-$\alpha$ lines is only about 2%. The joint probability that the clumped five Ly-$\alpha$ lines and the extensive metal line system would fall as close to each other as observed and not be physically associated is formally a very small number.

We suggest that the five clumped Ly-$\alpha$ lines plus the neighboring Ly-$\alpha$ line of the metal-line system are all part of the same complex, implying a total gaseous structure (containing Ly-$\alpha$ absorption lines and possibly galaxies) that extends over 3000 km s$^{-1}$ (20 Å). In recognition of this physical association, we have listed the properties of all six

Ly-$\alpha$ lines taken together in Table 17). This conjecture should be tested with GHRS (higher-resolution) spectra that would be expected to resolve clearly the Ly-$\alpha$ lines and the blended Ly-$\beta$ complex (cf. § 6.1.).

We note that there are no other clumps of Ly-$\alpha$ lines anywhere else in the observed spectrum of Ton 153 with a line density as high as the Ly-$\alpha$ clump near the extensive metal-line system. The next most impressive apparent clustering contains four Ly-$\alpha$ lines within a span of 16 Å ($z = 0.8294$, centered on 2223 Å).

Two clumps of Ly-$\alpha$ lines are present in the spectrum of PG 1352+011 near (in redshift space) extensive metal-line systems, $z = 0.6677$ and $z = 0.5258$. The characteristics of these two clumps are also given in Table 17. They contain, respectively, 4 Ly-$\alpha$ lines within a span of 9 Å and 5 Ly-$\alpha$ lines within 12 Å, including in both cases the Ly-$\alpha$ line of the extensive metal-line system. There are no other clumps of four or more Ly-$\alpha$ lines that lie within 20 Å of each other anywhere in the spectrum of PG 1352+011. If we ignore for the moment the Ly-$\alpha$ line of the metal-line system, The Press & Schechter (1974) formalism yields a probability of order 30% percent that one of the pure Ly-$\alpha$ clumps (3 Ly-$\alpha$ lines within 6 Å or 4 Ly-$\alpha$ lines within 9 Å) is the result of random coincidences. The average Ly-$\alpha$ line density in this spectrum is 0.0438 lines per Å. If we now take into account that in both cases an extensive metal-line system lies within less than 10 Å of the Ly-$\alpha$ clumps and that there are only two extensive metal-line systems in about 800 Å, we conclude that the Ly-$\alpha$ clumps and the neighboring metal-line systems are probably physically associated.

In summary, all three of the extensive metal-line systems detected in the spectra of Ton 153 and PG 1352+011 have neighboring clumps of Ly-$\alpha$ absorptions. The line density of Ly-$\alpha$ absorptions is enhanced by about an order of magnitude in the vicinity of the metal-line systems.

The evidence for Ly-$\alpha$ clumping around the three extensive metal-line systems found in the spectrum of PG 1634+706 is less strong. There are no Ly-$\alpha$ clumps detected around the $z$ = 1.045 system. However, there are 4 Ly-$\alpha$ lines in 15 Å around the $z$ = 0.9908 system and 4 Ly-$\alpha$ lines in 25 Å in the vicinity of the $z$ = 0.9060 system., including in both cases the Ly-$\alpha$ line of the neighboring metal-line system. Because of the relatively high line density of 0.0784 lines per Å in this spectrum, it would not be surprising to find by random coincidence one or two such clumps somewhere in the spectrum. In fact, there are other apparent Ly-$\alpha$ clumps in the spectrum of PG 1634+706, without detected neighboring metal-line systems, including 4 lines in 14 Å in the vicinity of $z$ = 1.10 and 4 lines within 14 Å near $z$ = 1.25.

It is somewhat suggestive of physical association that the $z$ = 0.9908 metal-line system lies within 9 Å of the neighboring clump of 3 pure Ly-$\alpha$ absorptions and the $z$ = 0.9060 metal-line system lies within 16 Å of its neighboring clump of 3 Ly-$\alpha$ lines. There are a total of 3 extensive metal-line systems in the approximately 210 Å of accessible spectrum.

In the spectrum of PKS 0122-00, which, because of a relatively poor signal-to-noise ratio, has the lowest average density of detected Ly-$\alpha$ absorption lines in the complete sample (0.023 lines per Å), there are no detected clumps of four or more Ly-$\alpha$ lines within 20 Å of the two extensive metal-line systems. However, in this case the two metal-line systems are very close to each other in redshift space, i. e., they are separated by only 2000 km s$^{-1}$. As argued in § 6.2., this pair of metal-line systems is probably physically associated since it is unlikely that the two most remarkable features in the spectrum of PKS 0122-00 should have–by chance–their Ly-$\alpha$ lines within 17 Å of each other.

The amount of Ly-$\alpha$ clumping that is detected depends upon, among other things, the resolution with which the spectra are observed, the method and the parameters used in selecting the absorption lines, and the rules adopted for identifications. It is hard to avoid some degree of *post facto* reasoning in the statistical considerations. Nevertheless, the overall pattern exhibited in Table 17 is impressive. The four best cases for clumping listed in Table 17 contain 19 Ly-$\alpha$ lines within 56 Å in the vicinity of the Ly-$\alpha$ lines of extensive metal-line systems. In an average 46 Å region in the spectra investigated here, only 2.5 Ly-$\alpha$ lines are present.

The velocity dispersions for these clumped associations of Ly-$\alpha$ lines are 600 km s$^{-1}$ to 1400 km s$^{-1}$ (see Table 17). The corresponding velocity spans, also listed in Table 17, are 1200 km s$^{-1}$ to 3000 km s$^{-1}$.

Figure 4 provides a clear visual representation of the clumping properties. This figure was constructed by performing a linear mapping, $X(\lambda)$, of the observed distribution of Ly-$\alpha$ lines in each spectrum onto the interval 0.0 to 1.0. Each line of wavelength $\lambda$ was plotted at the position

$$X(\lambda) = \frac{\lambda - \lambda_{\min}}{\lambda_{\max} - \lambda_{\min}} \quad , \qquad (7)$$

where $\lambda_{\max}$ is $(1 + z_{\mathrm{emission}}) \times 1215.67$ Å and $\lambda_{\min}$ is 10 Å less than the shortest wavelength at which a Ly-$\alpha$ line was observed in the complete spectrum. The Ly-$\alpha$ lines of the rare extensive metallic systems are shown as the longer vertical bars; the Ly-$\alpha$ lines without detected extensive metal-line systems are shown as the shorter vertical bars.

A quick visual inspection of the top panel (spectrum of Ton 153) of Figure 4 shows that the clump of six Ly-$\alpha$ lines at $X(\lambda) \simeq 0.3$, including the $z = 0.6606$ metal-line system, is remarkable. This clump is picked out by the eye as being unusual even when the extraordinary nature of the metallic system is not indicated, i.e., all Ly-$\alpha$ lines are plotted with the same vertical height. In the the third panel of Figure 4 (spectrum of PG 1352+011), the two clumps of five and of four Ly-$\alpha$ lines that include Ly-$\alpha$ lines of metal-line systems are also seen to be unusual. The possible clumping near two of the metal systems found in the spectrum of PG 1634+706 (see the bottom panel of Figure 4) seems to be, in agreement with the formal statistical calculations, not visually remarkable. For the three strongest cases of associated Ly-$\alpha$ clumping (near the $z = 0.6606$ system in the spectrum of Ton 153 and near the $z = 0.6677$ and the $z = 0.5258$ systems in the spectrum of PG 1352+011), the metallic Ly-$\alpha$ line lies on the boundary of the Ly-$\alpha$ clump (see the first and third panels of Figure 4). It is not clear if it is accidental that the Ly-$\alpha$ line from the extensive metal-line absorption system lies in all three cases near the boundary of the Ly-$\alpha$ clump.

The extraordinarily close association of the two extensive metal-line systems found in the spectrum of PKS 0122−00 is apparent in Figure 4b. There are only two such systems found anywhere in the spectrum of PKS 0122−00 and those two lie much closer together than the average separation of Ly-$\alpha$ lines from the more common Ly-$\alpha$ absorbers.

For future analysis, it is desirable to have a quantitative definition of what constitutes "a clump of Ly-$\alpha$ lines." We propose the following definition, which makes use of the definition of $X(\lambda)$ of Equation (7). A "clump" consists of three or more Ly-$\alpha$ lines which are separated from each other by less than 1/4 of the mean separation in X ($\sim 1/N$) for the sample. This definition requires that a clump have an overdensity of $\gtrsim 4$; the expected number of clumps can be readily calculated. This definition picks out the 3 clumps in the spectra of Ton 153 and PG 1352+011, but not the looser grouping in the spectrum of PG 16334+706.

We can calculate the product of the cross section and the spatial density of such systems by dividing the total numer of such systems by the total path length over which they would be visible in the spectra of the four quasars we have studied. Since there are several systematic uncertainties in this calculation, we estimate the path length in different ways. We define the accessible wavelength region for extensive metal-line systems as being the domain in which either 1) Si II $\lambda1190$ and C II $\lambda1335$ are both accessible; or in which 2) Si II $\lambda1190$ and C IV $\lambda1550$ are both accessible. For the three extensive metal line systems which show clear indications of being accompanied by Ly-$\alpha$ clumps we find

$$\left\langle \left(\frac{R}{10^{+2} \text{ kpc}}\right)^2 \left(\frac{n}{0.01 \text{ Mpc}^{-3}}\right) \right\rangle \simeq 4.24 \, h_{100}^{-1} \, z_{\text{path}}^{-1} \quad . \tag{8}$$

If we consider all eight of the extensive metal line systems tabulated in Table 17, then the value on the right hand of equation (8) becomes $11.3 \, h_{100}^{-1} \, z_{\text{path}}^{-1}$. Here $R$ is the characteristic radius of the clumps at zero redshift, $n$ is the number density of clumps, $h_{100}$ is the Hubble constant expressed in units of 100 km s$^{-1}$ Mpc$^{-1}$, and $z_{\text{path}}$ is the redshift path length.

We show in Table 18 the calculated column densities for several illustrative cases that span the plausible allowed range of parameter choices. The first four cases presented in Table 18 correspond to no cosmological evolution of the absorbing systems in either an $\Omega = 1$ or an $\Omega = 0.0$ universe; the results for the fifth and sixth cases are calculated assuming that the clumped systems evolve as rapidly as Ly-$\alpha$ clouds evolve at large redshift. For each choice of the cosmological model or for the assumed evolution of the Ly-$\alpha$ absorbers, the results are given for both assumptions about the required accessible wavelength region that is necessary to make possible the detection of an extensive metal-line system.

The column densities given in Table 18 are consistent with strucutres as common as galaxies but with extended halos of order $10^2$ kpc, or with structures as rare as rich clusters of galaxies but having larger radii of order 10 Mpc. The latter interpretation is more likely for those extensive metal line systems for which future studies confirm the reality of associated Ly-$\alpha$ clumps, since the velocity dispersions are larger than expected for single galaxies. We shall discuss in the concluding paragraphs of the following section some implications of the existence of clumped Ly−α absorbers near metal-line systems.

After completing the analysis of Ly-α clumping reported in this paper, we examined the lists of identifications of absorption line features that are reported in Paper I. The most extensive metal-line system that is described in Paper I, the $z = 0.7913$ system in the spectrum of PKS 2145+06, shows no evidence of a neighboring clump of Ly-α lines. There is evidence in the spectrum of 3C 351 for a modest clump of four Ly-α lines that includes the $z = 0.3646$ metal-line absorption system. The Ly-α line of the $z = 0.3646$ absorption system is itself very broad, $FWHM = 4.5$ Å. However, this 3C 351 system is a classic example of an 'associated absorption' complex, which lies only 1400 kms$^{-1}$ from the emission line redshift, and has been associated with an X-ray "warm" absorption system (Mathur et al. 1994). There is some evidence that some of these systems arise in outflows from the QSO. It is difficult to compare directly the clumping properties of Ly-α lines found in Paper I and in the present paper since the way that the line selection software, JASON, deals with blends has changed between the completion of Paper I and the present study. Nevertheless, it is appropriate to note that there is at least no contradiction between the results of the present study and the earlier investigation.

## 9. Summary And Discussion

We summarize and discuss in this section some results of the analysis presented in the preceding sections.

1. We have measured a total of 291 absorption lines in a complete sample with significance level greater than 4.5 in the FOS spectra of Ton 153, PKS 0122−00, PG 1352 +011, and PG 1634 + 706. The measurements were made using a set of computer algorithms whose effectiveness has been tested with the aid of Monte Carlo simulations. For each absorption line, the algorithms determine the observed wavelength and equivalent widths, the full-width at half-maximum, and their associated measurement uncertainties, as well as the statistical significance level (see § 3. and Paper II). The observed spectra and the equivalent width limits of sensitivity are shown in Figure 1.

2. We have identified the absorption lines using a set of computer algorithms ($ZSEARCH$) that are described in § 4. and in Paper I. The criteria for line identifications take account of atomic-physics constraints, the observed characteristics of the absorption spectra (including measurement errors and limits of sensitivity), the presence or absence of other expected "standard" absorption lines, the number of candidate absorption lines corresponding to a particular redshift, the strengths of candidate absorption lines relative to the strengths of Ly-$\alpha$, and, in a few specified cases, the relative strengths of metal lines from different ions. Potential multiple identifications of the same observed line at different candidate redshifts are resolved by a set of prioritized rules.

The probability of accidental redshift identifications is determined for C IV and O VI doublets by using the same software and the same identification procedures to search for pseudo-doublets in the real (observed) spectrum of each quasar. This Monte Carlo process preserves the complexities and the peculiarities of the observed spectra and of the identification algorithms. The results are discussed in § 5. and are summarized in Table 4 and Table 5; they were used as guides in making identifications, discussed in § 6., with real absorption doublets. The identification frequency for pseudo C IV doublets within the Ly-$\alpha$ forest varies by an order of magnitude among the different spectra considered in this paper and is significant for three of the four sources we discuss. The probability of finding pseudo C IV doublets outside the Ly-$\alpha$ forest is negligible. The probability is small of finding, anywhere in the observed spectra, pseudo O VI doublets with associated (real) Ly-$\alpha$ lines. With the aid of these simulations, identifications of C IV and O VI doublets can be made with confidence in the observed spectra.

3. For the four quasar spectra analyzed in this paper, the line selection software (JASON) found an average observed line density in the complete sample shortward of the Ly-$\alpha$ emission line that is 5.5 times larger than the average line density lines per Å, longward of Ly-$\alpha$ emission. This result is consistent with the interpretation (Lynds 1971), which is embodied in the line identification software (ZSEARCH), that most of the lines shortward of Ly-$\alpha$ emission are associated with hydrogenic absorption. Approximately 30% of the lines shortward of the Ly-$\alpha$ emission feature are identified metal-lines from extragalactic absorption systems or as Galactic interstellar lines. The remainder are identified as Ly-$\alpha$ absorption lines or higher members of the Lyman series. A total of 145 extragalactic Ly-$\alpha$ lines are identified in the complete sample of absorption lines measured in the four spectra analyzed here.

4. For $z_{\rm abs} \leq 1.3$, the density of Ly-$\alpha$ lines with rest equivalent widths greater than 0.24 Å is $(dN/dz)_0 = 24.3 \pm 6.6$ Ly-$\alpha$ lines per unit redshift, with $\gamma = 0.58 \pm 0.50$ (1-$\sigma$ uncertainties) and $dN/dz \propto (1+z)^\gamma$. These results are derived by applying to the total analyzed sample of Key Project quasar spectra (13 spectra discussed in Paper I, 4 spectra discussed in the present paper) a maximum likelihood estimator for the observed lines in those spectral regions in which the 4.5 $\sigma$ detection limit is less than 0.24 Å. The present results are in excellent agreement with previous determinations of the local line density of Ly-$\alpha$ lines using only small redshift quasars (see, e.g., Paper I and Bahcall et al. 1993b). The slope of the observed low-redshift $dN/dz$ relation differs at the $2 - 4.5\sigma$ level of significance from the slope deduced from various ground-based samples that refer to redshifts $z > 1.6$.

5. We identify a total of eight extensive metal-line systems, i.e., systems with four or more observed metal ion species, in the four moderate-redshift quasar spectra analyzed in the present paper. These systems are relatively rare: we find 0.0014 extensive metal-line systems per Å in the four spectra we analyze, whereas the Ly-$\alpha$ line density is 0.0453 per Å. For the extensive metal-line systems, $dN/dz \simeq 2.5(1 + z)^{0.5}$ systems per unit redshift or $dN/dz \simeq 2.0(1 + z)^{1.0}$ systems per unit redshift. The principal characteristics of the extensive metal-line systems are summarized in Table 17.

6. Two of the extensive metal systems, which are found in the spectrum of PKS 0122−00 at $z = 0.9667$ and $z = 0.9531$, are probably physically associated with each other since they are separated by only 2000 km s$^{-1}$; they appear to represent a manifestation of large scale structure that exists at the relatively early epoch of $z \sim 1$. We find only eight extensive metal-line systems in the approximately 2650 Å of accessible spectrum. Therefore, the *post facto* Poisson probability of finding two extensive metal-line systems as close as the two systems in PKS 0122−00, within 17 Å of each other, is less than 1%. For visual representation of the close association of these two unusual systems, see Figure 4. It is possible that these two systems are produced by a detectable cluster of galaxies or a supercluster of galaxies. Spectra of faint galaxies in this field are required in order to elucidate the dynamical properties of the absorption complex.

7. About half of the extensive metal-line systems are accompanied by highly-clustered, neighboring (in redshift space) clumps of Ly-$\alpha$ lines. There is strong statistical evidence, illustrated in Figure 4 and summarized in § 8. and in Table 17, that clumps of Ly-$\alpha$ lines are physically associated with the metal-line systems. For example, the joint random probability is less than 1% for finding the observed clump of Ly-$\alpha$ lines near the extensive metal-line system at $z = 0.6606$ in the spectrum of Ton 153. The probability is also small for finding in the spectrum of PG 1352+011 the conjunctions of metal-lines and Ly-$\alpha$ lines.

The line density of Ly-$\alpha$ absorptions is about an order of magnitude larger in the vicinity of several of the extensive metal-line systems than it is on the average in the spectra we have analyzed. If the statistical arguments presented in this paper are valid, then future observations of extensive metal-line systems at moderate redshifts should reveal Ly-$\alpha$ clumps in the vicinity of a significant fraction of the metal-line systems.

The local column density (number density times radius squared) for these clumps of Ly-$\alpha$ absorptions and metal-line systems is $10^{-4}$ Mpc$^{-1}$ (cf. § 8. and Table 18).

8. All four LLSs found in the spectra analyzed in this paper are identified with extensive metal-line systems (see Table 17). The LLS in Ton 153 (see Table 7) contains 10 lines in the ultraviolet spectra discussed here, all of which correspond to intermediate stages of ionization. The LLS with the most extensive metal-line counterpart appears in the spectrum of PG 1352+011 (see Table 12) and contains 18 lines from low, intermediate, and high-ionization stages. The spectrum of PG 1634+706 reveals two LLSs (see Table 15 and Table 16), containing 8 and 11 lines, respectively, from a wide range of ionization stages.

However, not all extensive metal-line systems correspond to a LLS. For example, the two strong metal-line systems in PKS 0122−00 that are shown in Table 9 and in Table 10 do not exhibit detectable Lyman-limit absorption ($\tau_{\rm LLS} < 0.25$). This result suggests that either the ionization is rather high or that the metal abundance must be relatively high for the two strong metal-line systems in PKS 0122−00 that do not show LLSs. This suggestion should be investigated with appropriate models for the ionization of the absorbing medium, together with higher resolution data.

In Paper I, we pointed out that the frequency of LLSs is consistent with their having sizes similar to the postulated sizes (Bahcall & Spitzer 1969) of gaseous galactic halos that could be responsible for metal-line absorption systems. The results obtained here, which show that LLSs are generally associated with strong metal-line absorption systems, provide further circumstantial evidence that LLSs arise predominantly in the extended halos of galaxies. This hypothesis could be tested by searching with ground-based observations for galaxies at the redshifts of the LLSs.

A systematic discussion of all the Key Project work to date on LLS is given by Stengler-Larrea et al. (1995).

9. There are no damped Ly-$\alpha$ candidates in the spectra presented in this paper. The total path length to detect damped Ly-$\alpha$ lines with rest equivalent widths greater than 10 Å in Key Project spectra is increased by approximately 10% by the observations analyzed in this paper. Our results continue to indicate that the number density per unit redshift of damped systems decreases with decreasing redshift (see Table 9, § 7, and conclusion 13 of Paper I).

Perhaps the most significant result of the present paper is the evidence, summarized in § 8. and in Table 17, that Ly-$\alpha$ lines are clustered near about half of the observed extensive metal-line systems. We suggest that the clumps of Ly-$\alpha$ lines and their accompanying metal-line systems, as well as the pair of neighboring metal-line systems found in the spectrum of PKS 0122+011, correspond to organized large structure at redshifts of between 0.5 and 1.0, analogous to clusters of galaxies or to superclusters observed previously at comparable or smaller redshifts. The velocity dispersions given in Table 17, 600 km s$^{-1}$ to 1400 km s$^{-1}$, are in agreement with the velocity range predicted by Bahcall & Salpeter (1966) for quasar absorption lines produced by clusters of galaxies and are similar to the observed velocity dispersions of clusters of galaxies. The observed column density of clumped systems, $10^{-4}$ Mpc$^{-1}$, is also consistent with what is predicted on the basis of the cluster-of-galaxy hypothesis, although the inferred cross sections are large compared to the cores of large clusters and thus may refer to cluster halos. The extensive metal-line systems may originate in either the halos of individual galaxies within a cluster of galaxies or within an inhomogeneous gaseous medium (assumed clumped to avoid predicting line widths that are larger than are observed) between the galaxies; the Ly-$\alpha$ absorptions may occur in the very extended halos of individual galaxies, or in gas associated with the cluster that has not yet formed galaxies, or has been tidal or ram pressure stripped. The many individual C IV absorption systems seen at large redshifts may be too numerous to be due to conventional clusters of galaxies.

The evidence presented in § 8. for clumping of Ly-$\alpha$ lines near extensive metal-line systems, together with the suggestive results presented in § 7. indicating that the rate of evolution of low redshift Ly-$\alpha$ lines is slower than the evolution found at larger redshifts, provides additional support for the suggestion (see Bahcall et al. 1992; Maloney 1992; Hoffman et al. 1993; Morris et al. 1993; Spinrad et al. 1993; Mo and Morris 1994) that there are at least two classes of Ly-$\alpha$ absorption lines: 1) the Ly-$\alpha$ lines, observed most easily at small redshifts, that are associated with galaxies and which, as shown in the present paper, can be clustered around extensive metal-line systems; and 2) the Ly-$\alpha$ lines that are not strongly clustered and which appear to dominate at large redshifts. According to this picture, the Ly-$\alpha$ lines of type 1 evolve at a rate similar to the evolution of Lyman-limit and metal-line systems, both of which are likely to be associated with galaxies. Type 2 systems evolve more rapidly. The present-day members of this type may be the very low column density absorbers that are not strongly associated with galaxies (cf. Morris et al. 1993). If this picture is correct, the two-point velocity correlation function of Ly-$\alpha$ lines may show significant power at splittings of a few hundred to a few thousand km s$^{-1}$ up to intermediate redshifts (i. e., up to $z \sim 1$ or up to $z \sim 1.5$) and, as has been previously demonstrated, very little power at these splittings at larger redshifts. It is possible that when the dynamical evolution of Ly-$\alpha$ absorbers is understood and when the variation of the ambient ionizing flux with redshift is determined, that some fraction of type 2 absorbers will be found to evolve into type 1 absorbers. If the evolution of one type into another type does occur, then the distinction between the two types of systems may become somewhat artificial. Discussions of the evolution of the number density of Ly-$\alpha$ absorbers without invoking separate populations may be found in, for example, Charlton, Salpeter, & Hogan (1993), Ikeuchi & Turner (1991), and Ikeuchi & Murakami (1994), but a detailed discussion of these models is beyond the scope of the present paper. The interested reader may also consult Bechtold (1995). Suggestive evidence for a change of slope in the number density of Ly-$\alpha$ absorption lines as a function of redshift, based on measurements of the average depression of quasar continua below the Ly-$\alpha$ emission lines, was presented by Schneider, Schmidt, & Gunn (1989) and is, within the uncertainties, consistent with the suggested evolution inferred here.

The evidence presented here for the existence of large-scale gaseous structures at redshifts of 0.5 to 1.0 with velocity dispersions of 600 km s$^{-1}$ to 1400 km s$^{-1}$ (or velocity spans of $12 \times 10^2$ km s$^{-1}$ to $3 \times 10^3$ km s$^{-1}$) provides a new constraint on cosmological models of structure formation.

This work was supported in part by NASA contracts NAG5-1618, HF-1045.01-93A, and grant GO-2424.01 from the Space Telescope Science Institute, which is operated by the Association of Universities for Research in Astronomy, Incorporated, under NASA contract NAS5-26555. D. Saxe wrote much of the Gaussian fitting software. We are grateful to W. Lane for valuable assistance with the simulations used to test the line finding and measurement software, and would like to thank Digital Equipment Corporation for providing the DEC4000 AXP Model 610 system used for the computationally-intensive parts of this project. Helpful comments on a draft copy of the manuscript by L. Lu are appreciated.

– 63 –Ford, H. C., & Hartig, G. F. 1990, Faint Object Spectrograph Instrument Handbook (Baltimore, MD: Space Telescope Science Institute)

Giallongo, E. 1991, MNRAS, 251, 541

Hoffman, G. L., Lu, N. Y., Salpeter, E. E., Farhat, B., Lamphier, C., & Roose, T. 1993, AJ, 106, 39

Hunstead, R. W. 1988, in QSO Absorption Lines, ed. J. C. Blades, D. Turnshek, & C. A. Norman (Cambridge, England: Cambridge University Press), 71

Ikeuchi, S., & Murakami, I. 1994, ApJ, 421, L79

Ikeuchi, S., & Turner, E. 1991, ApJ, 381, L1

Impey, C., Malkan, M., & collaborators 1994, private communication

Jannuzi, B. T., & Hartig, G. F. 1994, Calibrating Hubble Space Telescope, ed. J. C. Blades & S. J. Osmer, p. 215

Kirhakos, S., Sargent, W. L. W., Schneider, D. P., Bahcall, J. N., Jannuzi, B. T., Maoz, D., & Small, T. A. 1994, PASP, 106, 646

Lanzetta, K. M., Bowen, D. V., Tytler, D., & Webb, J. K. 1994, ApJ, 442, 538

Lasker, B. M., Conrad, R. S., McLean, B. J. Russell, J. L., Jenkner, H., & Shara, M. M. 1990, AJ, 99, 2019

Lockman, F. J., & Savage, B. D. 1995, ApJS, 97, 1

Lu, L. 1991, ApJ, 379, 99

Lu, L., & Savage, B. D. 1993, ApJ, 403, 127

Lu, L., Wolfe, A. M., & Turnshek, D. A. 1991, ApJ, 367, 19

Lynds, C. R. 1971, ApJ, 164 L73

# FIGURE CAPTIONS

FIG. 1.—Ultraviolet spectra of four quasars obtained with the Faint Object Spectrograph of the Hubble Space Telescope. Each panel contains a single spectrum obtained with the indicated grating (G270H, G190H, or G160L). The G270H and G190H observations were obtained with the 0.25″ × 2.0″ slit and the red digicon detector. The G160L observations were obtained using the blue digicon detector and the 1.0″ circular aperture. The short vertical bars indicate the positions of absorption lines in the complete sample. The dotted line is the "continuum" fit; see § 3. and Paper II. The lower line in each spectrum plot is the $1\sigma$ uncertainty in the flux as a function of wavelength. For the higher resolution observations, the $4.5\sigma_{\rm det}$ equivalent width limit (Å) for unresolved lines is also shown as a function of wavelength. Flat-field residuals that are strong enough that they would have a significance level greater than $4.5\sigma$ are marked with the symbol $FF$. For PG 1634+706, a minimum equivalent width limit of 0.135 Å was imposed, see § 3. To be included in the complete sample, a feature's equivalent width must exceed the value of the 4.5 $\sigma_{\rm det}$ curve at the relevant wavelength [see Eq. (2)]. The objects are presented in order of increasing redshift, as in the discussion in the text of the paper.

FIG. 2.—The normalized G270H grating FOS spectrum of PG 1634+706 is displayed together with the Gaussian fits made by the line finding and measurement software described in section 3. Fits are displayed for all lines with $SL > 3.0$. Only lines with $SL > 4.5$ are included in the complete sample described in the text and listed in the tables of line identifications.

FIG. 3.—The evolution of Ly-$\alpha$ absorption systems is shown over the redshift range from $z = 0.0$ to $z = 1.3$, based upon the Key Project quasars analyzed to date. The solid

line is the maximum-likelihood fit to a sample of 109 Ly-$\alpha$ lines with a uniform detection limit of 0.24 Å rest equivalent width. The best fit yields a value of $\gamma = 0.58 \pm 0.50$ assuming a distribution of the form $dN/dz \propto (1+z)^\gamma$. The larger, variable-detection-limit sample of 155 lines yields a value of $\gamma = 0.42 \pm 0.42$. For the uniform sample, we also show the local value of $(dN/dz)_z$ centered at each of four redshift bins of equal increments in $\Delta z$ between $z = 0.0$ and 1.3. The value of $(dN/dz)_z$ refers to the line density above a rest equivalent width of 0.24 Å. Including the set of all lines in our sample with rest equivalent widths greater than 0.24 Å, regardless of the detection limit at the wavelength in question, increases the sample size by about 50%. The maximum-likelihood estimation procedure described in Paper I yields for this variable-detection limit sample rather similar results: $(dN/dz)_0 = 25.5 \pm 5.3$ Ly-$\alpha$ absorption lines per unit redshift, with $\gamma = 0.42 \pm 0.42$. This distribution is shown above as the dashed line. Lines within 3000 km sec$^{-1}$ of the emission line redshift of all the quasars have been excluded from the samples analyzed in this figure.

FIG. 4.—The clumping properties of the Ly-$\alpha$ lines. For the four spectra, each Ly-$\alpha$ line, $\lambda$, is plotted at the position $X(\lambda) = (\lambda - \lambda_{\min})/(\lambda_{\max} - \lambda_{\min})$, where $\lambda_{\max} = (1 + z_{\text{emission}})\, 1215.67$ Å and $\lambda_{\min}$ is 10 Å less than the wavelength of the shortest wavelength Ly-$\alpha$ line in that spectrum. The vertical bars corresponding to the extensive metal-line systems are approximately twice as high as the vertical bars corresponding to the more common Ly-$\alpha$ lines.

– 68 –

Table 1. Journal of Observations

---

Table 2. Removed Flat-Field Features

---

Table 3. Zero-Point Offsets

---

Table 4. Pseudo C IV Doublets Found in Observed Spectra

---

Table 5. Pseudo O VI Doublets Found in Observed Spectra

---

Table 6. TON 153   $z=$ 1.0220

---

Table 7. The $z_{\mathrm{abs}} = 0.6606$ absorption system in TON 153

---

Table 8. PKS 0122-00  $z=$ 1.0700

Table 9. The $z_{\mathrm{abs}}$ = 0.9667 absorption system in PKS 0122-00

Table 10. The $z_{\mathrm{abs}}$ = 0.9531 absorption system in PKS 0122-00

Table 11. PG 1352+011  $z=$ 1.1210

Table 12. The $z_{\mathrm{abs}}$ = 0.6677 absorption system in PG 1352+011

Table 13. The $z_{\mathrm{abs}}$ = 0.5258 absorption system in the spectrum of PG 1352+011

Table 14. PG 1634+706  $z=$ 1.3340

Table 15. The $z_{\rm abs} = 1.0417$ absorption system (LLS) in the spectrum of PG 1634+706

Table 16. The $z_{\rm abs} = 0.9908$ absorption system in the spectrum of PG 1634+706

Table 17. Extensive Metal-Line Absorption System

Table 18. Column Density of Ly-$\alpha$ and Metal-line Clumps